\documentclass[final]{arxiv}
\usepackage{graphicx}
\usepackage{natbib}
\begin{document}

\title{Tidal Evolution of Close Binary Asteroid Systems}

\author{Patrick A. Taylor$^{1}$ and Jean-Luc Margot$^{2}$}
\affil{$^{1}$Arecibo Observatory, $^{2}$UCLA}
\email{ptaylor@naic.edu}

\journame{Celestial Mechanics and Dynamical Astronomy}
\submitted{27 February 2010}
\revised{12 August 2010}
\accepted{19 August 2010}
\pubonline{14 September 2010}
\pubprint{December 2010 in Volume 108, pp. 315--338}

\pages{40}
\tables{2}
\figures{8}

\qquad\qquad\qquad\qquad\qquad\qquad\,\,\,\,\,\copyright \, The Author(s) 2010.\\
\\
This article is published with open access at Springerlink.com and distributed under the terms of the
Creative Commons Attribution Noncommercial License which permits any noncommercial use, distribution,
and reproduction in any medium, provided the original author(s) and source are credited.\\
\\
Publisher's copy:  http://www.springerlink.com/content/735437q73h8v3372/fulltext.pdf\\
DOI address:  http://dx.doi.org/10.1007/s10569-010-9308-0
\clearpage


\noindent ABSTRACT:\\
\noindent
We provide a generalized discussion of tidal evolution to arbitrary order in the expansion of the
gravitational potential between two spherical bodies of any mass ratio.  To accurately reproduce the
tidal evolution of a system at separations less than five times the radius of the larger primary
component, the tidal potential due to the presence of a smaller secondary component is expanded in
terms of Legendre polynomials to arbitrary order rather than truncated at leading order as is typically
done in studies of well-separated system like the Earth and Moon.  The equations of tidal evolution
including tidal torques, the changes in spin rates of the components, and the change in semimajor axis
(orbital separation) are then derived for binary asteroid systems with circular and equatorial mutual
orbits.  Accounting for higher-order terms in the tidal potential serves to speed up the tidal evolution
of the system leading to underestimates in the time rates of change of the spin rates, semimajor axis,
and mean motion in the mutual orbit if such corrections are ignored.  Special attention is given to the
effect of close orbits on the calculation of material properties of the components, in terms of the
rigidity and tidal dissipation function, based on the tidal evolution of the system.  It is found that
accurate determinations of the physical parameters of the system, e.g., densities, sizes, and current 
separation, are typically more important than accounting for higher-order terms in the potential when 
calculating material properties.  In the scope of the long-term tidal evolution of the semimajor axis 
and the component spin rates, correcting for close orbits is a small effect, but for an instantaneous 
rate of change in spin rate, semimajor axis, or mean motion, the close-orbit correction can be on the 
order of tens of percent.  This work has possible implications for the determination of the Roche limit 
and for spin-state alteration during close flybys.\\
\\
\noindent
Keywords:  Gravity -- Extended Body Dynamics -- Tides -- Asteroids -- Binary asteroids
\pagebreak


\section{Introduction}
\label{sec:intro}

The classic equations for tidal evolution in two-body systems derived or utilized in seminal papers [e.g., 
\citet{macd64,gold66moon,gold66,mign79,mign81}], reviews [e.g., \citet{burn77,weid89,peal99}], and 
textbooks [e.g.,~\citet{murr99,danb92}] are based upon the underlying assumption that the two spherical 
components in the system are separated by several times the radius of the larger primary component.  While 
this assumption is valid in planet-satellite systems\footnote{There are small natural satellites of the 
outer planets that orbit very close to their primaries, but we must keep in mind that these satellites are 
part of much more complex dynamical systems than simple two-component binaries in addition to having 
negligible masses compared to their primaries.} such as Earth-Moon, Jupiter-Galilean satellites, and 
Saturn-Titan, as well as for Pluto-Charon and the majority of binary main-belt asteroids (with 100-km-scale 
primaries), it is not completely accurate for all binary asteroids, especially those in the near-Earth 
region.  Based upon the compilation by~\citet{wals06} of measured and estimated binary asteroid component 
size and semimajor axis parameters, nearly 75\% of near-Earth and Mars-crossing binaries have inter-component 
separations between 3 and 5 primary radii.  An updated compilation of parameters by~\citet{prav07} including 
small main-belt binaries, those with primaries less than 10~km in diameter, confirms that 75\% of binary 
systems among these three populations have close mutual orbits.  In addition, double asteroids, those 
systems with equal-size components that were not counted in the above tallies, such as (69230) 
Hermes~\citep{marg03,prav03,marg06iau}, (90) Antiope~\citep{merl00iauc,mich04,desc07}, (854) Frostia, (1089) 
Tama, (1313) Berna, and (4492) Debussy~\citep{behr06}, have separations within 5 primary radii.  The favored 
formation mechanism for near-Earth, Mars-crossing, and small main-belt binaries is rotational fission or 
mass shedding~\citep{marg02s,rich06,desc08} most likely due to YORP spin-up~\citep{prav07}, a torque on the 
asteroid spin state due to re-emission of absorbed sunlight~\citep{rubi00,vokr02}, where the typical 
binaries produced have equatorial mutual orbits with semimajor axes between 2 and 4.5 primary radii and 
eccentricities below 0.15~\citep{wals08yorp}.  Though all binary systems in these three populations may 
not have separations of less than 5 primary radii at present, if formed via spin-up, these systems likely 
have tidally evolved outward from a closer orbit.

Complex generalized formulae for tidal evolution are presented by~\citet{kaul64} and \citet{mign80} as 
extensions of the work of~\citet{darw79a,darw79b,darw80} that account for higher-order terms in the 
expansion of the tidal potential, though, nearly universally, even by Darwin, Kaula, and Mignard 
themselves, only the leading order is applied in practice under the assumption of a distant secondary 
and the negligibility of higher-order terms.  To date, the most common application of higher-order 
expansions of the tidal potential is in the Mars-Phobos system where tides on Mars raised by Phobos 
orbiting at 2.76 Mars radii are causing the gradual infall of Phobos's orbit.  As the separation between 
Mars and Phobos decreases, higher-order terms in the potential expansion must gain importance.  With 
this in mind, attempts to understand the observed secular acceleration of Phobos and the past history 
of its orbit date back to~\citet{redm64} and have continued with~\citet{smit76},~\citet{lamb79}, 
and~\citet{szet83}, among others, with~\citet{bill05} presenting the most recent treatment of the subject.

Because many binaries exist in a regime where traditional assumptions break down, and because tidal
evolution is most important at small separations, we are motivated to examine tidal interactions in 
close orbits.  Here, we expand the gravitational potential between two spherical bodies to arbitrary 
order as well as allow for a secondary of non-negligible mass.  We then present the resulting equations 
for the evolution of the component spin rates and the semimajor axis due to the tidal bulges raised on 
both components when restricted to systems with mutual orbits that are both circular and equatorial as 
suggested for small binaries formed via spin-up.  The effect of accounting for close orbits is examined 
and compared to the effect of uncertainties in physical parameters of the binary system.


\section{Tidal Potential of Arbitrary Order}
\label{sec:tidesl}

\begin{figure}[!ht]
\begin{center}
\includegraphics[angle=0., scale=0.8]{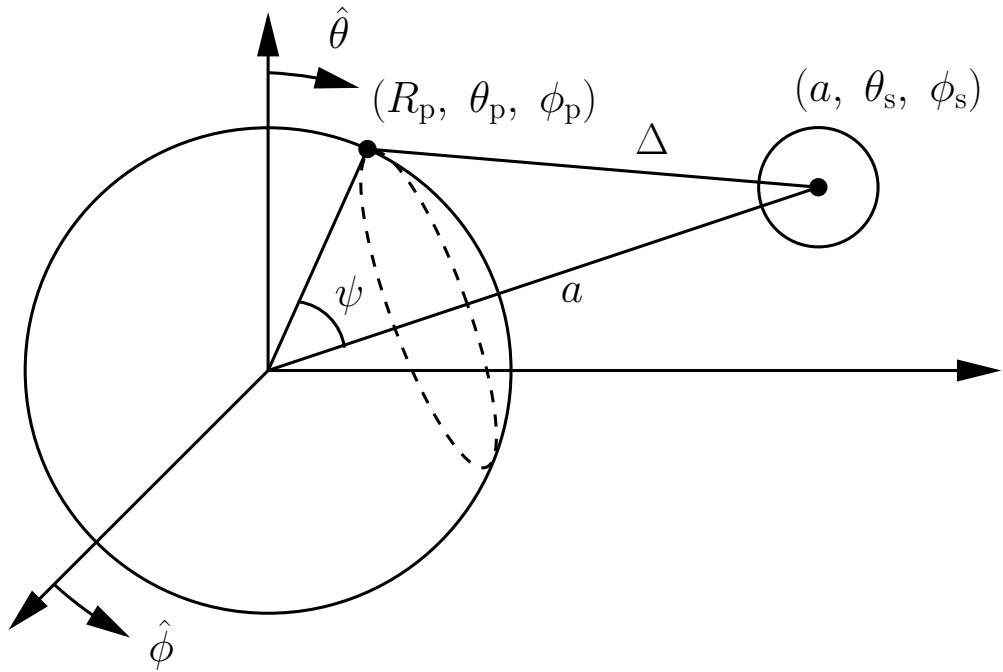}
\caption[]{Geometry for the potential felt on the surface of the primary due to the secondary orbiting a 
distance $a$ from the center of mass of the primary.  The $\textit{dashed line}$ is the locus of points on 
the surface of the spherical primary that are separated by the angle $\psi$ and distance $\Delta$ from the 
position of the spherical secondary and thus feel the same potential.}
\label{fig:sec2}
\end{center}
\end{figure}

The potential $V$ per unit mass at a point on the surface of the primary body of mass $M_{\rm p}$, radius 
$R_{\rm p}$, and uniform density $\rho_{\rm p}$ due to a secondary of mass $M_{\rm s}$, radius $R_{\rm s}$, 
and uniform density $\rho_{\rm s}$ orbiting on a prograde circular path in the equator plane of the 
primary with semimajor axis $a$ measured from the center of mass of the primary is

\begin{equation}
V=-G\,\frac{M_{\rm s}}{\Delta},
\label{eq:Vfull} 
\end{equation}

\noindent
where $G$ is the gravitational constant and $\Delta$ is the distance between the center of the secondary 
and the point of interest given by

\begin{equation}
\Delta=a\left[1-2\left(\frac{R_{\rm p}}{a}\right)\cos\psi+\left(\frac{R_{\rm p}}{a}\right)^2\right]^{1/2}, 
\end{equation} 

\noindent
with $\psi$ measured from the line joining the centers of the primary and secondary [e.g., \citet{murr99}].  
In the spherical polar coordinate system ($r$, $\theta$, $\phi$) shown in Fig.~\ref{fig:sec2}, with the 
polar angle $\theta$ measured from the rotation axis of the primary and the azimuthal angle $\phi$ measured 
from an arbitrary reference direction fixed in space, the separation angle $\psi$ between the secondary and 
the point of interest on the primary is

\begin{equation}
\cos\psi~=~\cos\theta_{\rm p}\cos\theta_{\rm s}+\sin\theta_{\rm p}\sin\theta_{\rm s}\cos\left(\phi_{\rm p}-\phi_{\rm s}\right).
\label{eq:psi}
\end{equation}  

\noindent
For widely separated binary systems where the semimajor axis $a$ is much larger than the radius of the 
primary $R_{\rm p}$, the potential is expanded in powers of the small term $R_{\rm p}/a$ such that

\begin{equation}
V=-G\,\frac{M_{\rm s}}{a}\left[1+\left(\frac{R_{\rm p}}{a}\right)\cos\psi+\left(\frac{R_{\rm p}}{a}\right)^2\,\frac{1}{2}\left(3\cos^2\psi-1\right)+\ldots\right].
\label{eq:Vexpansion}
\end{equation}

\noindent
The first term is independent of the position of the point of interest and thus produces no force on the 
primary.  The second term provides the force that keeps the mass element at the point of interest in a 
circular orbit about the center of mass of the system.  The third term is the tidal potential

\begin{equation}
U=-G\,\frac{M_{\rm s}R_{\rm p}^2}{a^3}\,\frac{1}{2}\left(3\cos^2\psi-1\right)
\label{eq:Ushort}
\end{equation}  

\noindent
that is the focus of past studies of tidal evolution where the the bodies are widely separated such as in 
the Earth-Moon system.  However, truncation of the expansion of $V$ in (\ref{eq:Vexpansion}) at three terms 
accurately estimates the true potential in (\ref{eq:Vfull}) only for separations exceeding 5$R_{\rm p}$.  
For smaller separations, as are often found among binary asteroids, higher orders in the expansion of $V$ 
are necessary.

The full expansion of the potential $V$ in~(\ref{eq:Vexpansion}) may be written concisely as the sum over 
Legendre polynomials $P_{\ell}(\cos\psi)$, i.e., zonal harmonic or azimuthally independent 
surface harmonic functions, as 

\begin{equation}
V=-G\,\frac{M_{\rm s}}{a}\,\sum_{\ell=0}^{\infty}\left(\frac{R_{\rm p}}{a}\right)^{\ell} P_{\ell}\left(\cos\psi\right),
\label{eq:Vlegendre}
\end{equation}

\noindent
where the $\ell=2$ term of the expansion of $V$ is the dominant tidal term in (\ref{eq:Ushort}).  The full 
tidal potential $U$ including all orders becomes

\begin{equation}
U=-G\,\frac{M_{\rm s}}{a}\,\sum_{\ell=2}^{\infty}\left(\frac{R_{\rm p}}{a}\right)^{\ell} P_{\ell}\left(\cos\psi\right).
\label{eq:Usum}
\end{equation}

\noindent
While we will derive the tidal evolution equations in terms of an arbitrary order $\ell$, 
Table~\ref{tab:legendre} lists the order $\ell$ of the expansion necessary for accurate reproduction of the 
potential $V$ at small separations.  At 2$R_{\rm p}$, the potential must be expanded to at least $\ell=6$, 
requiring four additional, but manageable, terms in the expansion.  This separation is convenient in terms 
of tidal evolution as it is the contact limit of a binary system with two equal-size components and a 
reasonable initial separation for the onset of tidal evolution in a newly formed binary system, regardless 
of component size, especially for systems formed through primary spin-up and mass shedding~\citep{wals08yorp}.  
Proceeding inward of 2$R_{\rm p}$ rapidly requires an unwieldy number of terms in the expansion 
(e.g., twice as many additional terms are needed at 1.5$R_{\rm p}$).


\section{Roche Limit}
\label{sec:roche}

The well-known classical fluid Roche limit is located at $a=2.46R_{\rm p}$~\citep{chan69} for equal 
density components, so that if one considers a secondary just outside the fluid limit, one must include 
the Legendre polynomials of orders $\ell\le4$ in the expansion for the potential felt by the primary.  
For solid, cohesionless\footnote{A cohesionless material has zero shear strength in the absence of 
confining pressure.  The interlocking of the constituent particles under pressure, however, can give the 
material shear strength.} bodies (gravitational aggregates or so-called rubble piles) modeled as a dry 
soil, the Roche limit falls approximately between $1.5R_{\rm p}$ and $2R_{\rm p}$~\citep{hols06,hols08,shar09}.  
The cohesionless Roche limit is based upon a binary system that is not tidally evolving, but the secondary 
remains stressed by its self-gravity, rotation (synchronized to the orbital period), and the difference in 
gravity from its near, primary-facing side to its far side.  \citet{hols06,hols08} illustrate that the mass 
ratio of the components has a negligible effect on the Roche limit, but one would expect that allowing the 
secondary to have a more rapid spin or allowing for higher-order tidal terms due to its proximity to the 
primary will increase the internal stresses and push the Roche limit farther from the primary, though, as 
noted by~\citet{shar09}, these issues have not been studied in detail.  

With a modest amount of cohesion, the secondary may exist within the stated Roche limit~\citep{hols08}.  
For the rough properties of a near-Earth binary of $\rho_{\rm p, s}=2$~g/cm$^{3}$ and $R_{\rm s}=100$~m, a 
cohesion value of $<100$~Pa is enough to hold the secondary together at the surface of the 
primary\footnote{The cohesion needed to prevent disruption scales as the square of both the density and 
size of the secondary.  Thus, for a main-belt binary with a $R_{\rm s}=10$~km, the necessary cohesion is of 
order 10$^{6}$~Pa, similar to monolithic rock.}.  For comparison, the surface material of comet Tempel 1 
excavated by the Deep Impact mission projectile is estimated to have a shear strength of 
$<65$~Pa~\citep{aher05} and an effective strength of 10$^{3}$~Pa~\citep{rich07}; fine-grained terrestrial 
sand is found to have cohesion values up to 250~Pa~\citep{schellart00}.  Therefore, it is not unreasonable 
that in the tidal field of the primary, the secondary can stably exist at the very least within the fluid 
Roche limit (even if cohesionless), if not also within the cohesionless Roche limit (with a cohesion 
comparable to comet regolith or sand), justifying our later choice to work to order $\ell=6$ corresponding 
to a separation of 2$R_{\rm p}$.  


\section{External Potential of Arbitrary Order}
\label{sec:extpotl}

The tidal potential $U_{\ell}$ of arbitrary order $\ell\ge2$ felt by the primary, taken from (\ref{eq:Usum}), 
may be written concisely as

\begin{equation}
U_\ell~=~-g_{\rm p}\,\zeta_{\ell, \rm p}\,P_{\ell}\left(\cos\psi\right),
\end{equation}

\noindent
where $g_{\rm p}=GM_{\rm p}/R_{\rm p}^2$ is the surface gravity of the primary and 

\begin{equation}
\zeta_{\ell, \rm p}~=~\frac{M_{\rm s}}{M_{\rm p}}\left(\frac{R_{\rm p}}{a}\right)^{\ell+1}R_{\rm p}.
\label{eq:zeta} 
\end{equation}

\noindent
The combination $\zeta_{\ell, \rm p}P_{\ell}\left(\cos\psi\right)$ is the equilibrium tide height, due
to the tidal potential of order $\ell$, that defines the equipotential surface about a primary that
is completely rigid (inflexible).  Because the mass ratio $M_{\rm s}/M_{\rm p}\le1$ and we assume 
$a\ge2R_{\rm p}$, the quantity $\zeta_{\ell, \rm p}/R_{\rm p}\le 1/8$ for all binary systems, and typically 
$\zeta_{\ell, \rm p}/R_{\rm p}\ll1$.  

For a body with realistic rigidity, the tidal potential $U_{\ell}$ physically deforms the surface of the 
primary by a small distance $\lambda_{\ell, \rm p}R_{\rm p}S_{\ell}$ as a function of position on the 
primary, where $\lambda_{\ell, \rm p}\ll1$ and $S_{\ell}$ is a surface harmonic function.  \citet{darw79a} 
and \citet{love27} lay the groundwork for showing that, in general, the deformation of a homogeneous 
density, incompressible sphere

\begin{equation}
\lambda_{\ell, \rm p}\,R_{\rm p}\,S_{\ell}~=~-\,h_{\ell, \rm p}\,\frac{U_{\ell}}{g_{\rm p}}~=~h_{\ell,\,\rm p}\,\zeta_{\ell,\,\rm p}\,P_{\ell}\left(\cos\psi\right)
\label{eq:lambdadef}
\end{equation}

\noindent
is given in terms of the displacement Love number $h_{\ell, \rm p}$~\citep{munk60},

\begin{equation}
h_{\ell, \rm p}~=~\frac{2\ell+1}{2\left(\ell-1\right)}\frac{1}{1+\frac{\left(2\ell^2+4\ell+3\right)\mu_{\rm p}}{\ell g_{\rm p}\rho_{\rm p}R_{\rm p}}},
\label{eq:hlove}
\end{equation}

\noindent
introducing $\mu_{\rm p}$ as the rigidity or shear modulus of the primary\footnote{\citet{darw79a} 
realized the correspondence between elastic and viscoelastic media and provides a generalized form for 
the deformation of a viscous spheroid, a function equivalent to~(\ref{eq:lambdadef}) he calls $\sigma$, 
that when applied to an elastic spheroid, in terms of rigidity rather than viscosity, is equivalent to 
the expression found here.}.  For bodies less than 200 km in radius, as all components of binary asteroid 
systems are, the rigidity $\mu$ dominates the stress due to self-gravity 
$g\rho R\sim G\rho^2R^2$~\citep{weid89}, even for rubble-pile structures (i.e., the model proposed 
by~\citet{gold09}), such that the Love number $h_{\ell}\ll1$ for small bodies.  With 
$h_{\ell, \rm p}$ and $\zeta_{\ell, \rm p}/R_{\rm p}$ small, and noting from~(\ref{eq:lambdadef}) that 
$\lambda_{\ell, \rm p}=h_{\ell, \rm p}\,\zeta_{\ell, \rm p}/R_{\rm p}$, the assumption of a small 
deformation factor $\lambda_{\ell, \rm p}$ is justified.     

Of particular interest is the external potential felt by the secondary now that the primary has 
been deformed.  It is this external potential that produces the tidal torque that transfers
angular momentum through the system.  Here, we slightly alter our spherical coordinate system such 
that $\theta$ now measures the angle from the axis of symmetry of the tidal bulge, as in~\citet{murr99}, 
such that the surface of the nearly spherical primary is now given by 

\begin{equation}
R~=~R_{\rm p}\left(1+\sum_{\ell=2}^{\infty}\lambda_{\ell, \rm p} P_{\ell}\left(\cos\theta\right)\right).
\label{eq:shell}
\end{equation}

\noindent
The potential felt at a point external to the primary is the sum of the potential of a spherical 
primary with radius $R_{\rm p}$ and that of the deformed shell.  However, only that due to the deformed 
shell, called the non-central potential by~\citet{murr99}, will contribute to the torque.

\begin{figure}[!ht]
\begin{center}
\includegraphics[angle=0., scale=0.8]{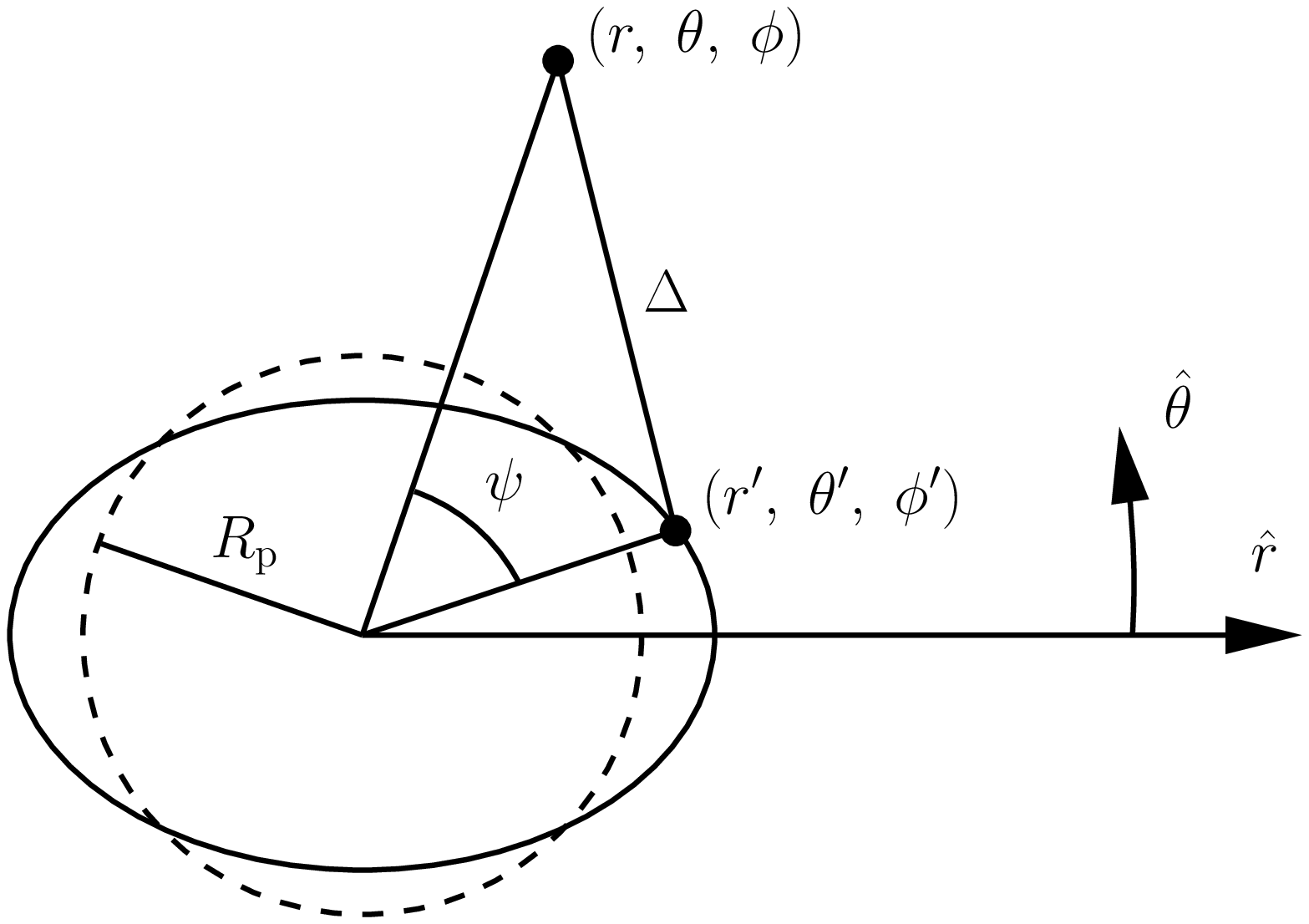}
\caption[]{Geometry for the potential felt at an external point due to the deformation of the primary from 
its initially $\textit{spherical shape}$ ($dashed$).  Note that here $\theta$ is measured from the axis of 
symmetry of the tidal bulge.}
\label{fig:sec4}
\end{center}
\end{figure}

In Fig.~\ref{fig:sec4}, the reciprocal of the distance $\Delta$ between the external point 
($r$, $\theta$, $\phi$) and a point on the surface of the primary 
($r^{\prime}$, $\theta^{\prime}$, $\phi^{\prime}$) separated by an angle $\psi$, where 
$r^{\prime}=R$ from~(\ref{eq:shell}), is 

\begin{equation}
\frac{1}{\Delta}~=~\frac{1}{r}\sum_{\ell=0}^{\infty}\left(\frac{R_{\rm p}}{r}\right)^{\ell}P_{\ell}\left(\cos\psi\right)~+~O\left(\lambda_{\ell^{\prime}, \rm p}\right).
\label{eq:delta}
\end{equation} 

\noindent
The use of $\ell^{\prime}$ denotes terms based upon the surface deformation rather than the expansion 
of the distance between the points of interest.  The non-central potential (per unit mass of the 
object disturbed by the potential) due to the deformed shell with mass element 
$\displaystyle{\rho_{\rm p}R_{\rm p}^3\sum_{\ell^{\prime}=2}^{\infty}\lambda_{\ell^{\prime}, \rm p}P_{\ell^{\prime}}\left(\cos\theta^{\prime}\right)d\left(\cos\theta^{\prime}\right)d\phi^{\prime}}$ is

\begin{equation}
U_{\rm nc}~=~-G\rho_{\rm p} R_{\rm p}^2\left(\frac{R_{\rm p}}{r}\right)\sum_{\ell^{\prime}=2}^{\infty}\sum_{\ell=0}^{\infty}\lambda_{\ell^{\prime}, \rm p}\left(\frac{R_{\rm p}}{r}\right)^{\ell}\int\int P_{\ell^{\prime}}\left(\cos\theta^{\prime}\right)P_{\ell}\left(\cos\psi\right)d\left(\cos\theta^{\prime}\right)d\phi^{\prime},
\end{equation}

\noindent
where the double integral goes over the surface of the primary.  The integral of the product of
two surface harmonics like the Legendre polynomials over a surface is zero unless $\ell=\ell^{\prime}$
such that for a specific order $\ell\ge2$~\citep{macr67}, 

\begin{eqnarray}
U_{\ell, \rm nc} & = & -G\rho_{\rm p} R_{\rm p}^2\left(\frac{R_{\rm p}}{r}\right)\lambda_{\ell, \rm p}\,\times\,\frac{4\pi}{2\ell+1}\left(\frac{R_{\rm p}}{r}\right)^{\ell}P_{\ell}\left(\cos\theta\right)\nonumber\\
 & = & -\frac{3}{2\ell+1}h_{\ell, \rm p}\zeta_{\ell, \rm p}g_{\rm p}\left(\frac{R_{\rm p}}{r}\right)^{\ell+1}P_{\ell}\left(\cos\theta\right).
\end{eqnarray}

\noindent
By defining the more familiar potential Love number

\begin{equation}
k_{\ell, \rm p}~=~\frac{3}{2\ell+1}h_{\ell, \rm p}~=~\frac{3}{2\left(\ell-1\right)}\frac{1}{1+\frac{\left(2\ell^2+4\ell+3\right)\mu_{\rm p}}{\ell g_{\rm p}\rho_{\rm p}R_{\rm p}}},
\label{eq:klove}
\end{equation}

\noindent
which is of a similar order as $h_{\ell, \rm p}$, the non-central potential is written in the form

\begin{equation}
U_{\ell, \rm nc}~=~-k_{\ell, \rm p}g_{\rm p}\zeta_{\ell, \rm p}\left(\frac{R_{\rm p}}{r}\right)^{\ell+1}P_{\ell}\left(\cos\theta\right)
\label{eq:Uexternal}
\end{equation}

\noindent
such that $U_{\ell, \rm nc}$ at the surface of the primary is simply $k_{\ell, \rm p}U_{\ell}$.  
Because $\mu \gg g\rho R$ for small bodies, the Love number $k_{\ell, \rm p}$ may be approximated by

\begin{equation}
k_{\ell, \rm p}~\simeq~\frac{3}{2\left(\ell-1\right)}\,\frac{\ell}{2\ell^2+4\ell+3}\frac{g_{\rm p}\rho_{\rm p}R_{\rm p}}{\mu_{\rm p}}~=~\frac{2\pi}{\ell-1}\,\frac{\ell}{2\ell^2+4\ell+3}\frac{G\rho_{\rm p}^2 R_{\rm p}^2}{\mu_{\rm p}}.
\label{eq:kapprox}
\end{equation}

\noindent
Taking the external point to be the position of the secondary orbiting at a distance $a$ from the 
primary, the complete\footnote{Here, by complete we mean accounting for all orders $\ell$.  We have, 
however, limited the result to first order in the Love number $k_{\ell, \rm p}$ because terms of order 
$\lambda_{\ell, \rm p}$ were ignored in (\ref{eq:delta}).  These would have produced higher-order terms 
in the Love number in the final form of the potential in (\ref{eq:Ufull}), but because we have argued 
$\lambda_{\ell, \rm p}$ and $k_{\ell, \rm p}$ are both small quantities, terms of second and higher 
order in the Love number are negligible.} non-central potential per unit secondary mass due to tides 
raised on the primary is 

\begin{eqnarray}
U_{\rm nc} & = & -g_{\rm p}\sum_{\ell=2}^{\infty}k_{\ell, \rm p}\zeta_{\ell, \rm p}\left(\frac{a}{R_{\rm p}}\right)^{-\left(\ell+1\right)}P_{\ell}\left(\cos\theta\right) \nonumber \\
 & = & -\frac{GM_{\rm s}}{R_{\rm p}}\sum_{\ell=2}^{\infty}k_{\ell, \rm p}\left(\frac{a}{R_{\rm p}}\right)^{-2\left(\ell+1\right)}P_{\ell}\left(\cos\theta\right).
\label{eq:Ufull}
\end{eqnarray}

\noindent
The non-central potential drops off quickly with increasing separation as the separation to the sixth
power for $\ell=2$ and by an additional square of the separation for each successive order.  The 
$\theta$ term in the Legendre polynomial accounts for the angular separation between the external
point of interest and the tidal bulge of the primary.  For the specific location of the secondary, we
define the angle $\delta$ as the geometric lag angle between the axis of symmetry of the tidal 
bulge and the line connecting the centers of the two components.  


\section{Tidal Dissipation Function $Q$}
\label{sec:Q}

In addition to the rigidity $\mu$, the response of a homogeneous, incompressible sphere to a disturbing 
potential is characterized by the tidal dissipation function $Q$ defined by 

\begin{equation}
Q^{-1}~=~\frac{1}{2\pi E^{*}}\oint \left(-\frac{dE}{dt}\right)dt,
\end{equation}

\noindent where $E^{*}$ is the maximum energy stored in the tidal distortion and the integral is the 
energy dissipated over one cycle [see~\citet{gold63} or~\citet{efro09} for detailed discussions].  This 
definition is akin to the quality factor in a damped, linear oscillator and does not depend on the 
details of how the energy is dissipated.  Friction in the response of the body to a tide-raising 
potential plus the rotation of the body itself (at a spin rate $\omega$ compared to the mean motion $n$ 
in the mutual orbit about the center of mass of the system) lead to misalignment by the geometric lag 
angle $\delta$.  

The geometric lag relates to a phase lag by $\epsilon_{\ell m p q}=-m\,\delta\,{\rm sign}\left(\omega-n\right)$, 
where the $\ell m p q$ notation follows~\citet{kaul64}, and we have implicitly assumed a single tidal 
bulge as done by~\citet{gers55} and~\citet{macd64} by using a single positive geometric lag $\delta$ 
independent of the tidal frequencies\footnote{The definition of the phase lag~\citep{kaul64,efro09}, 
when one ignores changes in the periapse and node, is 
$\epsilon_{\ell m p q}= \left[\left(\ell-2p+q\right)n-m\omega\right]\Delta t_{\ell m p q}$, where the 
bracketed term is the tidal frequency and $\Delta t_{\ell m p q}$ is the positive time lag in the response
of the material to the tidal potential.  In the potential expansion by~\citet{kaul64}, only terms satisfying 
$\ell-2p=m$ and $q=0$ survive for mutual orbits that are circular and equatorial such that 
$\epsilon_{\ell m p q}=-m\,|\omega-n|\,\Delta t_{\ell m p q}\,{\rm sign}\left(\omega-n\right)=-m\,\delta\,{\rm sign}\left(\omega-n\right)$, 
assuming a constant time lag and a single (positive) value for geometric lag for all viable combinations 
of ${\ell m p q}$.}.  The tidal dissipation function $Q$, in turn, relates to the phase angle as 
$Q^{-1}_{\ell m p q}=|{\rm cot}\,\,\epsilon_{\ell m p q}| \simeq |\epsilon_{\ell m p q}| + O(\epsilon_{\ell m p q}^{2})$~\citep{efro09} 
provided energy dissipation is weak ($Q_{\ell m p q} \gg 1$).  The absolute value of 
$\epsilon_{\rm \ell m p q}$ is required on physical grounds to ensure that $Q_{\ell m p q}$ is positive.  
Since the tidal dissipation function is related to the phase lag, a different $Q_{\ell m p q}$ 
technically applies to each tidal frequency.  Compared to the dominant order $\ell=2$, where only the 
${\ell m p q}=2200$ term survives in the setup of our problem,

\begin{equation}
Q_{\ell m p q}^{-1}=m\delta=\frac{m}{2}\,Q_{2200}^{-1}=\frac{m}{2}\,Q^{-1}
\label{eq:Q}
\end{equation}

\noindent
in general, where we define $Q \equiv Q_{2200}$ such that $Q_{\ell m p q}$ for any tidal frequency is
proportional to a single value of $Q$.  This simple relation between $Q_{\ell m p q}$ and the $Q$ of the
dominant tidal frequency is a direct result of our assumption of a single geometric lag independent of 
tidal frequency.  Such a choice may not be the most realistic physical model\footnote{In our model, $Q$ 
varies inversely with the tidal frequency.  \citet{efro09} argue in favor of a rheological model where
$Q$ scales to a positive fractional power of the tidal frequency (at least for terrestrial planets).  It 
is unclear what rheological model is proper for gravitational aggregates like binary asteroids.}, but 
does allow for simpler mathematical manipulation.  Because $Q$ is necessarily positive regardless of the 
sign of the phase lag, we append ${\rm sign}\left(\omega-n\right)$ to our forthcoming equations, where 
the spin rate $\omega$ relates to the tidally distorted component.  If $\omega > n$, the bulge leads; if 
$\omega < n$, the bulge lags behind.


\section{Tidal Torques on the Components}
\label{sec:torques}

\begin{figure}[!ht]
\begin{center}
\includegraphics[angle=0., scale=0.8]{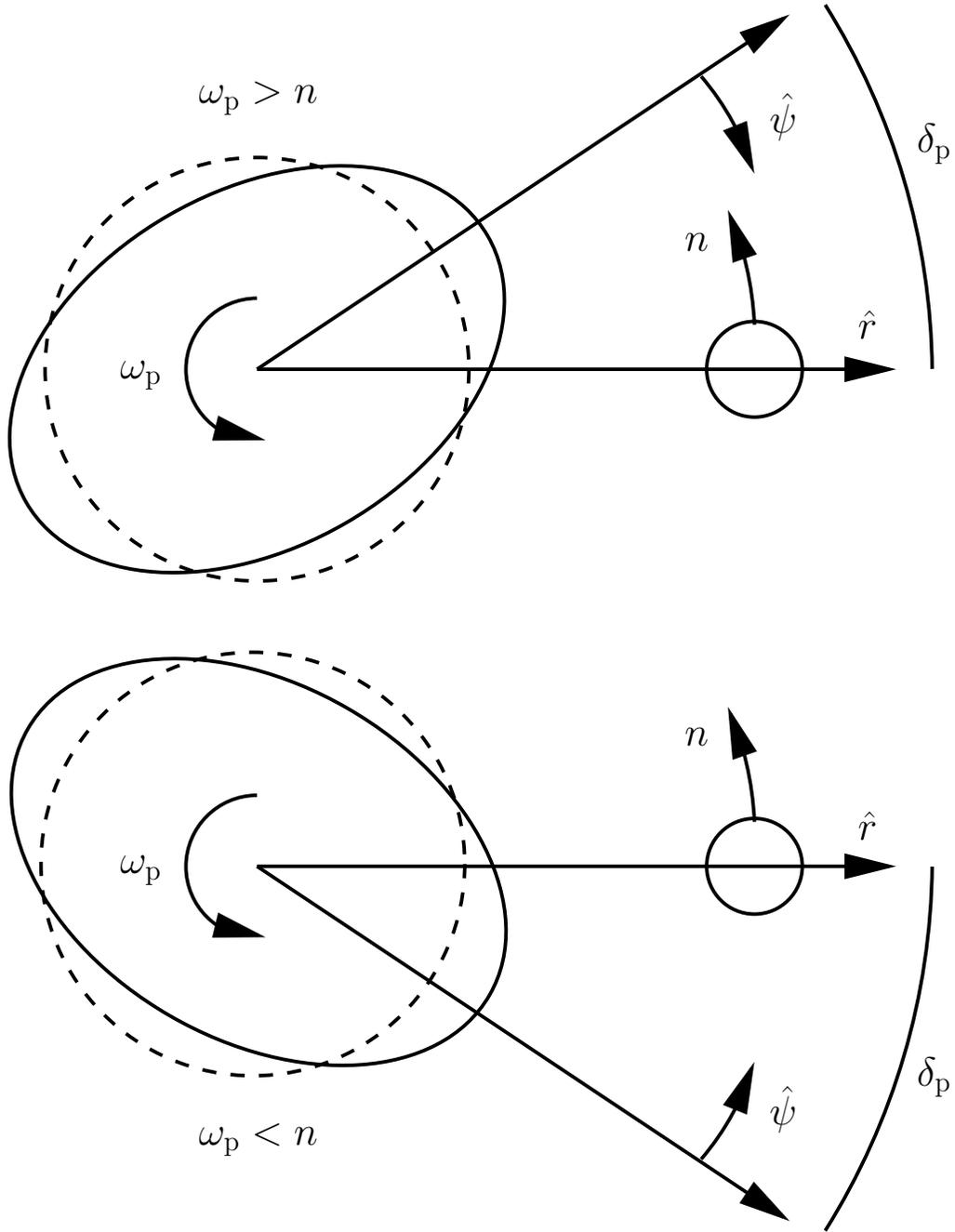}
\caption[]{When the primary spins faster than the mean motion of the mutual orbit $\left(\omega_{\rm p}>n\right)$, 
the tidal bulge is carried ahead of the tide-raising secondary.  The resulting torques slow the rotation 
of the primary and expand the mutual orbit.  When the primary spins slower than the mean motion 
$\left(\omega_{\rm p}<n\right)$, the torques speed up the rotation of the primary and contract the mutual 
orbit.  Similar diagrams apply to tides raised on the secondary and whether or not $\omega_{\rm s}>n$.  
Note that $\psi$ is measured from the axis of symmetry of the tidal bulge of the primary with 
$\psi=\delta_{\rm p}$ being the geometric lag angle at the position of the secondary.}
\label{fig:sec6}
\end{center}
\end{figure}

The force on the secondary due to the distorted primary at order $\ell$ is $-M_{\rm s}\nabla U_{\ell, \rm nc}$, 
and because we have restricted the problem to a circular, equatorial mutual orbit, the tidal bulge 
remains in the orbit plane, and the sole component of the force is tangential to the mutual orbit.  
Returning to the notation where $\psi$ measures the angle from the axis of symmetry of the tidal 
bulge\footnote{In this notation, the tidal potential in~(\ref{eq:Usum}) deforms the shape of the component 
according to~(\ref{eq:lambdadef}) and produces the external potential~(\ref{eq:Ufull}) all in terms of the 
single angle $\psi$.}, the force at the location of the secondary is proportional to 
$\displaystyle{-\left.\partial P_{\ell}/\partial\psi\right|_{\psi=\delta}}$ and pointed in the 
$-\hat{\psi}$ direction.  The value of $\delta$ is taken to be positive as stated in the previous section
such that, for $\delta$ small, the quantity $\displaystyle{-\left.\partial P_{\ell}/\partial\psi\right|_{\psi=\delta}}$ 
is positive and the primary bulge attracts the secondary.  For a prograde mutual orbit with 
$\omega_{\rm p}>n$, the primary bulge pulls the secondary ahead in the orbit; if $\omega_{\rm p}<n$, the 
primary bulge retards the motion of the secondary (see Fig.~\ref{fig:sec6}).  The resulting torque vector 
acting upon the orbit of the secondary, which is located at position $\textbf{r}$ with respect to the 
center of mass of the primary, is given by 
\textbf{$\Gamma_{\ell, \rm p}$}$=\textbf{r}\times\left(-M_{\rm s}~\nabla U_{\ell, \rm nc}\right)$.  Thus, 
the torque vector, in general, is proportional to 
$\displaystyle{-\left.\partial P_{\ell}/\partial\psi\right|_{\psi=\delta}\,\left(\hat{\psi}\times\hat{r}\right)}$.  
As defined, the direction (sign) of $\hat{\psi}\times\hat{r}$ depends on whether the tidal bulge leads or 
lags, and we indicate this in the magnitude of the torque via the term ${\rm sign}\left(\omega-n\right)$
such that the torque on the orbit of the secondary due to the $\ell^{\rm th}$-order deformation of the 
primary is

\begin{eqnarray}
\Gamma_{\ell, \rm p} & = & -M_{\rm s}\frac{\partial U_{\ell, \rm nc}}{\partial \psi_{\rm p}}\nonumber\\
 & = & k_{\ell, \rm p}\frac{GM_{\rm s}^2}{R_{\rm p}}\left(\frac{a}{R_{\rm p}}\right)^{-2\left(\ell+1\right)}\left(\left.-\frac{\partial P_{\ell}\left(\cos\psi_{\rm p}\right)}{\partial \psi_{\rm p}}\right|_{\psi_{\rm p}=\delta_{\rm p}}\right){\rm sign}\left(\omega_{\rm p}-n\right).
\label{eq:torque}
\end{eqnarray}

\noindent
where $\delta_{\rm p}$ is the geometric lag angle between the primary's tidal bulge and the line 
of centers.  A positive (negative) torque increases (decreases) the energy of the orbit at a rate 
$\Gamma_{\rm p}\,n$.  An equal and opposite torque alters the rotational energy of the primary at a rate 
$-\Gamma_{\rm p} \omega_{\rm p}$ such that the total energy $E$ of the system is dissipated over time at 
a rate $\dot{E}=-\Gamma_{\rm p}\left(\omega_{\rm p}-n\right)<0$ as heat inside the primary.  Though energy 
is dissipated, angular momentum is conserved due to the equal and opposite nature of the torques on the 
orbit and the rotation of the primary.  Conservation of angular momentum results in the evolution of the 
mutual orbit and is discussed in the following section.

A similar torque arises from tides raised on the secondary.  By the symmetry of motion about the center 
of mass, the torque $\Gamma_{\ell, \rm s}$ is given by swapping the subscripts p and s in (\ref{eq:torque}) 
such that

\begin{eqnarray}
\Gamma_{\ell, \rm s} & = & k_{\ell, \rm s}\frac{GM_{\rm p}^2}{R_{\rm s}}\left(\frac{a}{R_{\rm s}}\right)^{-2\left(\ell+1\right)}\left(-\left.\frac{\partial P_{\ell}\left(\cos\psi_{\rm s}\right)}{\partial\psi_{\rm s}}\right|_{\psi_{\rm s}=\delta_{\rm s}}\right)\,{\rm sign}\left(\omega_{\rm s}-n\right)\label{eq:torques}\\
 & = & k_{\ell, \rm p}\frac{GM_{\rm s}^2}{R_{\rm p}}\left(\frac{R_{\rm s}}{R_{\rm p}}\right)^{2\ell-3}\frac{\mu_{\rm p}}{\mu_{\rm s}}\left(\frac{a}{R_{\rm p}}\right)^{-2\left(\ell+1\right)}\left(-\left.\frac{\partial P_{\ell}\left(\cos\psi_{\rm s}\right)}{\partial\psi_{\rm s}}\right|_{\psi_{\rm s}=\delta_{\rm s}}\right)\,{\rm sign}\left(\omega_{\rm s}-n\right),\nonumber
\end{eqnarray}

\noindent
where $\delta_{\rm s}$ is the geometric lag angle between the tidal bulge of the secondary and the line
of centers.  This torque changes the orbital energy at a rate $\Gamma_{\rm s}\,n$, and the equal and 
opposite torque alters the rotational energy of the secondary at a rate $-\Gamma_{\rm s}\omega_{\rm s}$, 
dissipating energy as heat in the secondary at a rate $\dot{E}=-\Gamma_{\rm s}\left(\omega_{\rm s}-n\right)$.  
Torques on the primary and secondary weaken for higher orders of $\ell$ and increasing separations, as 
expected, and do so in the same manner as the non-central potential in (\ref{eq:Ufull}).  Once the 
rotation rate of a component synchronizes with the mean motion of the mutual orbit, the associated torque 
goes to zero due to the ${\rm sign}\left(\omega-n\right)$ term\footnote{If the mutual orbit were not 
circular, a radial tide owing to the eccentricity would continue to act despite the synchronization of 
the component spin rate to the mean motion.}.  Note that we have ignored interactions between the tidal 
bulges of the components as these will depend on the square (or higher powers) of the Love numbers, which 
we have argued are negligible (see Footnote 5).


\section{Spin Rate and Semimajor Axis Evolution for Close Orbits}
\label{sec:eqns}

During tidal evolution, angular momentum is transferred between the spins of the components and the 
mutual orbit.  For simplicity, assume that the primary and secondary have spin axes parallel to the 
normal of the mutual orbit plane and rotate in a prograde sense.  Then, the torque on the distorted 
primary alters its spin with time at a rate $\dot{\omega_{\rm p}}=-\Gamma_{\rm p}/I_{\rm p}$, where 
$I_{\rm p}=\alpha_{\rm p}M_{\rm p}R_{\rm p}^2$ is the moment of inertia of the primary.  The pre-factor 
$\alpha$ is $2/5$ for a uniform density sphere, but can vary with the internal structure of the body, 
and is left as a variable here such that the change in spin rate of the primary is  

\begin{equation}
\dot{\omega}_{\ell, \rm p}~=~-\frac{k_{\ell, \rm p}}{\alpha_{\rm p}}\,\frac{\kappa^2}{1+\kappa}\left(\frac{a}{R_{\rm p}}\right)^{-2\ell+1}n^2\left(-\left.\frac{\partial P_{\ell}\left(\cos\psi_{\rm p}\right)}{\partial \psi_{\rm p}}\right|_{\psi_{\rm p}=\delta_{\rm p}}\right){\rm sign}\left(\omega_{\rm p}-n\right),
\label{eq:wp}
\end{equation}

\noindent
recalling that $-\partial P_{\ell}/\partial\psi\ge0$ for small angles and defining the mass ratio 
$\kappa \equiv M_{\rm s}/M_{\rm p}=\left(\rho_{\rm s}/\rho_{\rm p}\right)\left(R_{\rm s}/R_{\rm p}\right)^3$.  
Also note that $n^2$, which is proportional to $\left(a/R_{\rm p}\right)^{-3}$, was introduced via 
Kepler's Third Law, $n^2a^3=G\left(M_{\rm p}+M_{\rm s}\right)$.  For rapidly spinning primaries with 
$\omega_{\rm p}>n$, the torque will slow the rotation.  

To conserve angular momentum in the system, the change in spin angular momentum, given by the torque 
$-\Gamma_{\ell, \rm p}$, plus the change in orbital angular momentum must be zero.  The orbital 
angular momentum for a circular mutual orbit $M_{\rm p}M_{\rm s}/\left(M_{\rm p}+M_{\rm s}\right)\,na^{2}$
changes with time as $\left(1/2\right) M_{\rm p}M_{\rm s}/\left(M_{\rm p}+M_{\rm s}\right)\,na\dot{a}$
such that conservation requires

\begin{equation}
\left(\frac{\dot{a}}{R_{\rm p}}\right)_{\ell, \rm p}~=~2k_{\ell, \rm p}\,\kappa\left(\frac{a}{R_{\rm p}}\right)^{-2\ell}n\left.\left(-\frac{\partial P_{\ell}\left(\cos\psi_{\rm p}\right)}{\partial\psi_{\rm p}}\right|_{\psi_{\rm p}=\delta_{\rm p}}\right) {\rm sign}\left(\omega_{\rm p}-n\right)
\label{eq:adotp}
\end{equation}

\noindent
for each order $\ell$.  For rapidly spinning primaries, the orbit will expand as angular momentum is 
transferred from the spin of the primary to the mutual orbit and, so long as the geometric lag remains 
small, higher orders will cause both more rapid despinning of the primary and faster expansion of the 
mutual orbit than $\ell=2$ alone.  A large secondary with $\kappa\sim1$ clearly causes the most rapid 
tidal evolution.  A small secondary with $\kappa\ll1$ will not cause the primary to despin appreciably 
due to the $\kappa^2$-dependence of (\ref{eq:wp}), but the separation will evolve more readily as 
(\ref{eq:adotp}) scales as $\kappa$.  

One can derive the change in the semimajor axis in (\ref{eq:adotp}) by other methods including the work 
done on the orbit and Gauss's formulation of Lagrange's planetary equations.  Setting the time derivative 
of the total energy of the orbit $-GM_{\rm p}M_{\rm s}/2a$, which is $GM_{\rm p}M_{\rm s}\dot{a}/2a^{2}$, 
equal to the work done on the orbit $\Gamma_{\ell, \rm p}n$ simplifies to (\ref{eq:adotp}).  Using Gauss's 
formulation for spherical bodies (see \citet{burn76} for a lucid derivation) and a circular mutual orbit, 

\begin{equation}
\dot{a}_{\ell}~=~\frac{2}{n}\left(1+\kappa\right)T_{\ell},
\label{eq:gauss}
\end{equation}

\noindent
where $T_{\ell}$ is the tangential component of the disturbing force (per unit mass) from the previous 
section, which is $\left(1/a\right)\partial U_{\ell, \rm nc}/\partial \psi~{\rm sign}\left(\omega_{\rm p}-n\right)$ 
with $U_{\ell, \rm nc}$ given by (\ref{eq:Uexternal}).  The $\left(1+\kappa\right)$ term is not typically 
present in the Gauss formulation, but is appended here to the disturbing function\footnote{Algebraically, 
from the time rate of change of the orbital energy, $\displaystyle{\dot{a}=2a^{2}\dot{E}/GM_{\rm p}M_{\rm s}}$, 
and the change in orbital energy is further related to the velocity of the secondary ${\bf \dot{r}}$ and the 
disturbing force ${\bf F}=-M_{\rm s}\nabla U_{\rm nc}$ such that $\dot{E}={\bf \dot{r}}\cdot{\bf F}=naM_{\rm s}T$ 
for a circular mutual orbit.  Replacing $\dot{E}$ by $naM_{\rm s}T$ and using Kepler's Third Law, 
$n^{2}a^{3}=G\left(M_{\rm p}+M_{\rm s}\right)=GM_{\rm p}\left(1+\kappa\right)$, in the expression for $\dot{a}$ 
gives (\ref{eq:gauss}) for a specific order $\ell$.  If $M_{\rm s}$ were ignored in Kepler's Third Law, the 
more familiar form of Gauss's formulation would emerge:  $\dot{a}=2T/n$.} due to the non-inertial nature of 
the coordinate system centered on the primary~\citep{rubi73} and is necessary, as stated in~\citet{darw80}, 
because the primary ``must be reduced to rest.''  One can also argue the term is necessary to account for 
the reaction of one body to the tidal action of the other~\citep{ferr08} as the perturbation is internal to 
the binary system rather than an external element (e.g., drag force, third body).  The $\left(1+\kappa\right)$ 
term is absent in the formulae of~\citet{kaul64}, which is reasonable if the secondary is of negligible mass, 
but we wish to allow for an arbitrary mass ratio.  Substitution of the disturbing force into (\ref{eq:gauss}) 
produces (\ref{eq:adotp}).  By including the $\left(1+\kappa\right)$ term in Kaula's equation (38), evaluating 
Kaula's $F$ and $G$ functions with zero inclination and eccentricity, and recalling that Kaula's 
$\epsilon_{\ell m p q} = -m\,\delta\,{\rm sign}\left(\omega_{\rm p}-n\right)$ in our notation, we find our 
evolution of the semimajor axis in (\ref{eq:adotp}) is a special case of Kaula's generalization\footnote{The 
product of Kaula's $F_{\ell m p}$ and $G_{\ell p q}$ functions is non-zero for circular, equatorial orbits 
only if $\ell-2p=m$ and $q=0$.  The prefactors on each $\psi$ in the Legendre polynomials listed in 
Table~\ref{tab:legendre} are the values of $m$ for each order $\ell$ that satisfy $\ell-2p=m$.  Thus, the 
cosine terms in the Legendre polynomials we list correspond to $\ell m p q$ of 2200, 3110, 3300, 4210, 4400, 
5120, 5310, 5500, 6220, 6410, and 6600.  This correspondence allows us to link our equations written in terms 
of Legendre polynomials and a geometric lag to Kaula's equations written in terms of the phase lag 
$\epsilon_{\ell m p q}$.  While the combinations 2010, 4020, and 6030 satisfy $\ell-2p=m$, terms with $m=0$ 
cannot contribute to the tidal evolution of the system because, by definition, these terms do not produce a 
phase lag.  These three terms are responsible for the $\psi$-independent components of the Legendre polynomials 
with $\ell=2,4,6$ that vanish upon differentiation with respect to $\psi$.}, as one would expect.

The tidal evolution of the secondary follows similarly.  The torque on the distorted secondary alters its 
spin with time at a rate $\dot{\omega_{\rm s}}=-\Gamma_{\rm s}/I_{\rm s}$,

\begin{eqnarray}
\dot{\omega}_{\ell, \rm s} & = & -\frac{k_{\ell, \rm s}}{\alpha_{\rm s}}\frac{1}{\kappa\left(1+\kappa\right)}\left(\frac{R_{\rm s}}{R_{\rm p}}\right)^{2\ell-1}\left(\frac{a}{R_{\rm p}}\right)^{-2\ell+1}n^2\left(-\left.\frac{\partial P_{\ell}\left(\cos\psi_{\rm s}\right)}{\partial \psi_{\rm s}}\right|_{\psi_{\rm s}=\delta_{\rm s}}\right){\rm sign}\left(\omega_{\rm s}-n\right)\label{eq:ws}\\
 & = & -\frac{k_{\ell, \rm p}}{\alpha_{\rm p}}\frac{\kappa}{1+\kappa}\left(\frac{R_{\rm s}}{R_{\rm p}}\right)^{2\ell-5}\frac{\alpha_{\rm p}}{\alpha_{\rm s}}\,\frac{\mu_{\rm p}}{\mu_{\rm s}}\,\left(\frac{a}{R_{\rm p}}\right)^{-2\ell+1}n^2\,\left(-\left.\frac{\partial P_{\ell}\left(\cos\psi_{\rm s}\right)}{\partial \psi_{\rm s}}\right|_{\psi_{\rm s}=\delta_{\rm s}}\right){\rm sign}\left(\omega_{\rm s}-n\right),\nonumber
\end{eqnarray} 

\noindent
and alters the semimajor axis of the mutual orbit at a rate of

\begin{equation}
\left(\frac{\dot{a}}{R_{\rm p}}\right)_{\ell, \rm s}~=~2k_{\ell, \rm p}\,\kappa\left(\frac{R_{\rm s}}{R_{\rm p}}\right)^{2\ell-3}\,\frac{\mu_{\rm p}}{\mu_{\rm s}}\,\left(\frac{a}{R_{\rm p}}\right)^{-2\ell}n\,\left.\left(-\frac{\partial P_{\ell}\left(\cos\psi_{\rm s}\right)}{\partial\psi_{\rm s}}\right|_{\psi_{\rm s}=\delta_{\rm s}}\right) {\rm sign}\left(\omega_{\rm s}-n\right).
\label{eq:adots}
\end{equation}

\noindent
The Legendre polynomials in Table~\ref{tab:legendre} are written as sums of terms of the form $\cos\,m\psi$ 
where $m$ is an integer (see Footnote 11).  Thus, the derivative 
$\displaystyle{\left.\partial P_{\ell}/\partial\psi\right|_{\psi=\delta}}$ is a sum of terms of the form 
$\sin\,m\delta$.  For small geometric lag angles $(Q \gg 1)$, 
$\displaystyle{\left.-\partial P_{\ell}/\partial\psi\right|_{\psi=\delta} \ge 0}$ and 
$\sin\,m\delta \simeq m\delta$ such that 
$\displaystyle{\left.-\partial P_{\ell}/\partial\psi\right|_{\psi=\delta}}$ 
$\displaystyle{\propto Q^{-1}}$.  Because the derivative of a Legendre polynomial is proportional to $Q^{-1}$, 
only the size ratio of the components and their material properties, in terms of their respective $\mu Q$ 
values, determine the relative strength of the torques and the relative contributions to the orbit expansion,

\begin{equation}
\left|\frac{\Gamma_{\ell, \rm s}}{\Gamma_{\ell, \rm p}}\right|~=~\left|\frac{\dot{a}_{\ell, \rm s}}{\dot{a}_{\ell, \rm p}}\right|~=~\left(\frac{R_{\rm s}}{R_{\rm p}}\right)^{2\ell-3}\,\frac{\mu_{\rm p}Q_{\rm p}}{\mu_{\rm s}Q_{\rm s}},
\label{eq:aratio}
\end{equation}

\noindent
with the relative contribution of the secondary decreasing at higher orders of $\ell$ and for smaller 
secondaries.  Note that the relative strength of the torques is independent of the mass and 
density\footnote{However, the absolute strengths of the torques in (\ref{eq:torque}) and (\ref{eq:torques}) 
do depend on the masses and densities of the components.}.  For classical $\ell=2$ tides on components with 
similar material properties, the torque due to the distorted secondary is a factor of the size ratio 
weaker than the torque due to the distorted primary.  For each higher order in the expansion, the relative 
strength of the torque due to the distorted secondary weakens by the square of the size ratio.  The 
changes in the spin rates compare as

\begin{equation}
\left|\frac{\dot{\omega}_{\ell, \rm s}}{\dot{\omega}_{\ell, \rm p}}\right|~=~\frac{1}{\kappa}\left(\frac{R_{\rm s}}{R_{\rm p}}\right)^{2\ell-5}\,\frac{\alpha_{\rm p}}{\alpha_{\rm s}}\,\frac{\mu_{\rm p}Q_{\rm p}}{\mu_{\rm s}Q_{\rm s}}~=~\frac{\rho_{\rm p}}{\rho_{\rm s}}\left(\frac{R_{\rm s}}{R_{\rm p}}\right)^{2(\ell-4)}\,\frac{\alpha_{\rm p}}{\alpha_{\rm s}}\,\frac{\mu_{\rm p}Q_{\rm p}}{\mu_{\rm s}Q_{\rm s}}.
\label{eq:wratio}
\end{equation}

\noindent
This differs from a generalization of Darwin's result [c.f. \citet{darw79b}, p. 521] because we 
have included the ratio of the Love numbers of the components.  At the dominant orders, $\ell=2$ and $3$, 
with similar densities, shapes, and material properties, the spin rate of the secondary changes faster 
than the primary.  However, interestingly, for $\ell=4$, the contributions to the changes in spin rates are 
equal, and for orders $\ell>4$, the contribution to the change in spin rate of the primary is greater than 
that of the secondary.  As with the torques, the relative strength of the changes in spin rates weakens by 
the square of the size ratio for each successive order $\ell$.  For smaller secondaries, the changes in 
spin rates are smaller than for similar mass components, and, for all cases, the process of changing the 
spin of the primary is slower than for the secondary.

Evaluating the Love number $k_{\ell, \rm p}$ in (\ref{eq:kapprox}) and $\partial P_{\ell}/\partial\psi_{\rm p}$ 
from Table~\ref{tab:legendre} explicitly for orders $\ell\le6$, assuming a small geometric lag angle 
$\delta_{\rm p}$, and applying (\ref{eq:Q}), the spin of the primary changes as 

\begin{eqnarray}
\dot{\omega}_{\rm p} & = & -\frac{8}{19}\frac{1}{\alpha_{\rm p}}\frac{\pi^2 G^2\rho_{\rm p}^{3}R_{\rm p}^2}{\mu_{\rm p}Q_{\rm p}}\,\kappa^2\left(\frac{a}{R_{\rm p}}\right)^{-6}\,{\rm sign}\left(\omega_{\rm p}-n\right)\nonumber\\
 & & \quad \times\left[1+\frac{19}{22}\left(\frac{a}{R_{\rm p}}\right)^{-2}+\frac{380}{459}\left(\frac{a}{R_{\rm p}}\right)^{-4}+\frac{475}{584}\left(\frac{a}{R_{\rm p}}\right)^{-6}+\frac{133}{165}\left(\frac{a}{R_{\rm p}}\right)^{-8}\right],
\label{eq:wpall}
\end{eqnarray}

\noindent 
where $n$ has been replaced with Kepler's Third Law to show the full dependence upon the separation
of the components $a/R_{\rm p}$.  Using either (\ref{eq:ws}) or (\ref{eq:wratio}), the spin of the 
secondary changes as

\begin{eqnarray}
\dot{\omega}_{\rm s} & = & -\frac{8}{19}\frac{1}{\alpha_{\rm s}}\frac{\pi^2 G^2\rho_{\rm p}^{3}R_{\rm p}^2}{\mu_{\rm s}Q_{\rm s}}\,\kappa\left(\frac{R_{\rm s}}{R_{\rm p}}\right)^{-1}\left(\frac{a}{R_{\rm p}}\right)^{-6}\,{\rm sign}\left(\omega_{\rm s}-n\right)\nonumber\\
 & & \quad \times\left[1+\frac{19}{22}\left(\frac{R_{\rm s}}{R_{\rm p}}\right)^2\left(\frac{a}{R_{\rm p}}\right)^{-2}+\frac{380}{459}\left(\frac{R_{\rm s}}{R_{\rm p}}\right)^4\left(\frac{a}{R_{\rm p}}\right)^{-4}\right.\nonumber\\
 & & \quad \quad \left.+\frac{475}{584}\left(\frac{R_{\rm s}}{R_{\rm p}}\right)^6\left(\frac{a}{R_{\rm p}}\right)^{-6}+\frac{133}{165}\left(\frac{R_{\rm s}}{R_{\rm p}}\right)^8\left(\frac{a}{R_{\rm p}}\right)^{-8}\right].
\label{eq:wsall}
\end{eqnarray}

\noindent
Assuming similar densities for the components, the change in the spin rate of the primary scales as the
size ratio of the components to the sixth power ($\propto \kappa^{2}$);  the spin rate of the secondary 
scales only as the square of the size ratio at leading order, reinforcing from (\ref{eq:wratio}) how the 
spin of the secondary evolves more rapidly than that of the primary, especially for small size ratios.  

For close orbits, the separation of the components changes as angular momentum is transferred to 
or from the spins of the components such that the overall change in the orbital separation for 
$\ell \le 6$ is the sum of (\ref{eq:adotp}) and (\ref{eq:adots}),

\begin{eqnarray}
\frac{\dot{a}}{R_{\rm p}} & = & \frac{8\sqrt{3}}{19}\frac{\pi^{3/2}G^{3/2}\rho_{\rm p}^{5/2}R_{\rm p}^2}{\mu_{\rm p}Q_{\rm p}}\,\kappa\left(1+\kappa\right)^{1/2}\left(\frac{a}{R_{\rm p}}\right)^{-11/2}\nonumber\\
 & & \enspace \times\left[{\rm sign}\left(\omega_{\rm p}-n\right)+\left(\frac{R_{\rm s}}{R_{\rm p}}\right)\frac{\mu_{\rm p}Q_{\rm p}}{\mu_{\rm s}Q_{\rm s}}{\rm sign}\left(\omega_{\rm s}-n\right)\right.\nonumber\\
 & & \enspace +\frac{19}{22}\left(\frac{a}{R_{\rm p}}\right)^{-2}\left({\rm sign}\left(\omega_{\rm p}-n\right)+\left(\frac{R_{\rm s}}{R_{\rm p}}\right)^3\frac{\mu_{\rm p}Q_{\rm p}}{\mu_{\rm s}Q_{\rm s}}{\rm sign}\left(\omega_{\rm s}-n\right)\right)\nonumber\\
 & & \enspace +\frac{380}{459}\left(\frac{a}{R_{\rm p}}\right)^{-4}\left({\rm sign}\left(\omega_{\rm p}-n\right)+\left(\frac{R_{\rm s}}{R_{\rm p}}\right)^5\frac{\mu_{\rm p}Q_{\rm p}}{\mu_{\rm s}Q_{\rm s}}{\rm sign}\left(\omega_{\rm s}-n\right)\right)\nonumber\\
 & & \enspace +\frac{475}{584}\left(\frac{a}{R_{\rm p}}\right)^{-6}\left({\rm sign}\left(\omega_{\rm p}-n\right)+\left(\frac{R_{\rm s}}{R_{\rm p}}\right)^7\frac{\mu_{\rm p}Q_{\rm p}}{\mu_{\rm s}Q_{\rm s}}{\rm sign}\left(\omega_{\rm s}-n\right)\right)\nonumber\\
 & & \enspace \left.+\frac{133}{165}\left(\frac{a}{R_{\rm p}}\right)^{-8}\left({\rm sign}\left(\omega_{\rm p}-n\right)+\left(\frac{R_{\rm s}}{R_{\rm p}}\right)^9\frac{\mu_{\rm p}Q_{\rm p}}{\mu_{\rm s}Q_{\rm s}}{\rm sign}\left(\omega_{\rm s}-n\right)\right)\right].
\label{eq:adotboth}
\end{eqnarray}

\noindent
Inside the square brackets, having a secondary of negligible size ($R_{\rm s}/R_{\rm p}\rightarrow0$) 
has the same effect as having a synchronous secondary ($\omega_{\rm s}=n$); both make the contribution
from the secondary vanish.  Of course, if one considers the factor outside the square brackets, having a 
secondary of negligible size makes the mass ratio $\kappa$ negligible, while having a synchronous secondary 
does not directly affect $\kappa$.  The change in the mean motion of the mutual orbit follows from Kepler's 
Third Law and (\ref{eq:adotboth}) as

\begin{equation}
\frac{\dot{n}}{n}~=~-\frac{3}{2}\left(\frac{a}{R_{\rm p}}\right)^{-1}\left(\frac{\dot{a}}{R_{\rm p}}\right).
\label{eq:ndot}
\end{equation}

\noindent
Note that in the above equations (\ref{eq:wpall}--\ref{eq:ndot}), any difference in density between the 
components is accounted for in the mass ratio $\kappa$; otherwise, only the size ratio of the components is 
involved in the terms due to tides raised on the secondary.  Obviously, the contribution of the secondary 
is most important when the components are of similar size.  Not only is the contribution of the secondary 
weakened because of its smaller size, it should also be despun faster than the primary such that its 
contribution turns off when $\omega_{\rm s}=n$ long before the primary does the same.  Furthermore, each 
equation has a strong inverse dependence on the separation of the components even at $\ell=2$, and while
the inclusion of higher-order terms will be strongest at small separations, the orbit of a typical 
outwardly evolving system will expand to a wider separation rapidly. 


\section{Effect of Close Orbit Expansion on Tidal Evolution}

\begin{figure}[!ht]
\begin{center}
\includegraphics[angle=0., scale=0.8]{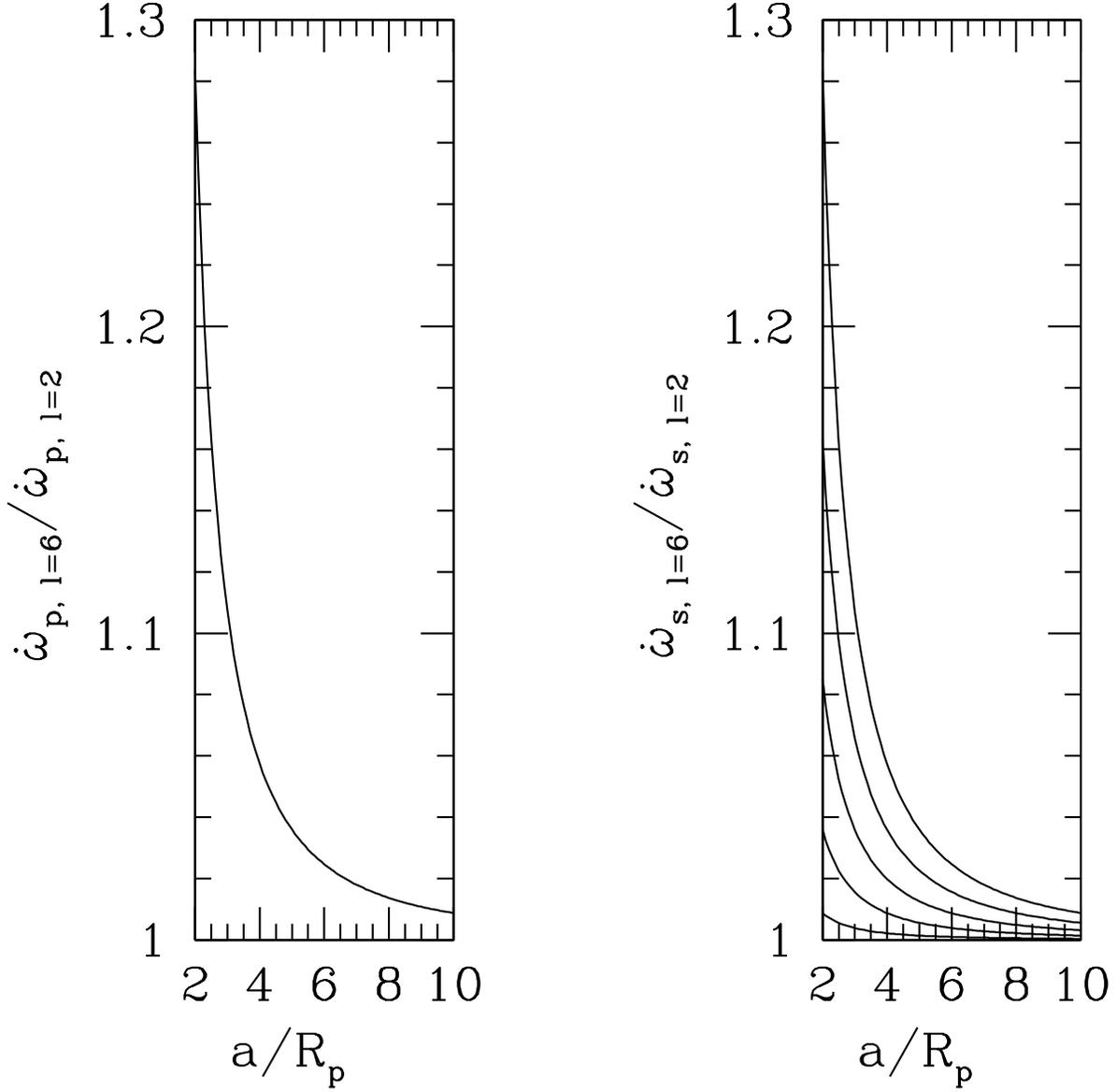}
\caption[]{Time rates of change of the spin rate of the primary ($left$) and secondary ($right$) as a
function of the separation of the components using all orders up to $\ell=6$ versus using classical 
$\ell=2$ tides only.  The plotted ratio amounts to the $\textit{bracketed portions}$ of (\ref{eq:wpall}) 
and (\ref{eq:wsall}) for the primary and secondary, respectively.  The change in spin rate of the 
primary due to higher-order terms is unaffected by the size ratio of the components.  The change in 
spin rate of the secondary is greater for larger size ratios, plotted from $top$ to $bottom$ with 
$R_{\rm s}/R_{\rm p}=1, 0.8, 0.6, 0.4,$ and $0.2$.  The effect of higher-order terms is always 
below 30\% beyond 2$R_{\rm p}$ and falls below 1\% by a separation of 10$R_{\rm p}$ for all size 
ratios.}
\label{fig:wdot}
\end{center}
\end{figure}

\begin{figure}[!ht]
\begin{center}
\includegraphics[angle=0., scale=0.8]{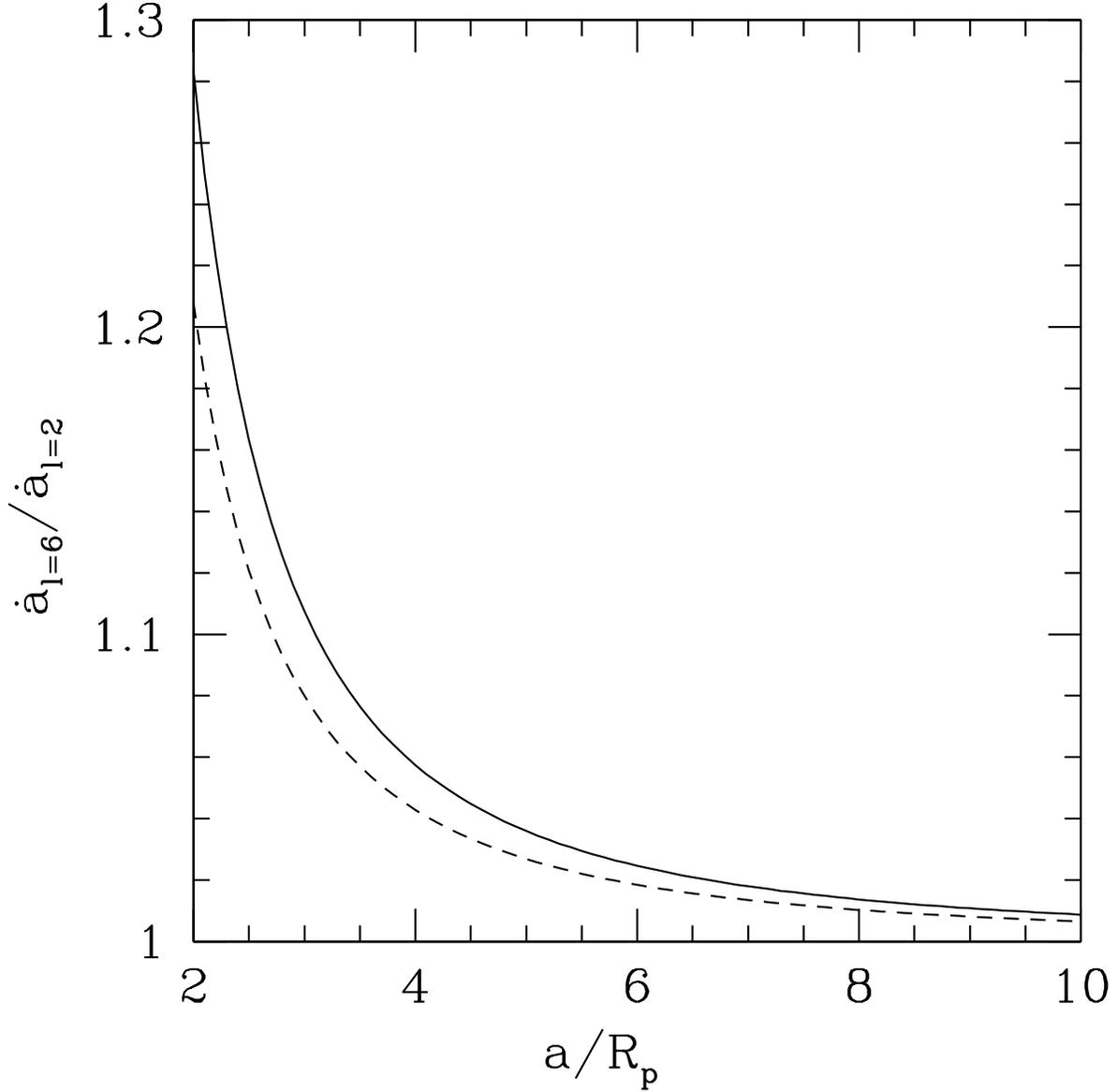}
\caption[]{Time rate of change of the semimajor axis of the mutual orbit as a function of the separation 
of the components using all orders up to $\ell=6$ versus using classical $\ell=2$ tides.  The plotted 
ratio amounts to the $\textit{bracketed portion}$ of (\ref{eq:adotboth}) divided by $1+R_{\rm s}/R_{\rm p}$ 
with both components having similar $\mu Q$ parameters and contributing to the evolution in an additive 
sense.  The $\textit{solid curve}$ corresponds to a system with components of equal size ($R_{\rm s}/R_{\rm p}=1$), 
a secondary of negligible size ($R_{\rm s}/R_{\rm p}=0$), or a synchronized secondary ($\omega_{\rm s}=n$).  
The lower bound ($\textit{dashed curve}$) is for the size ratio $R_{\rm s}/R_{\rm p}=0.53$.  As in 
Fig.~\ref{fig:wdot}, the effect of higher-order terms is always below 30\% beyond 2$R_{\rm p}$ and falls 
below 1\% by a separation of 10$R_{\rm p}$ for all size ratios.}
\label{fig:adot}
\end{center}
\end{figure}

Inclusion of higher-order terms for the changes in spin rates and semimajor axis in 
(\ref{eq:wpall}--\ref{eq:adotboth}) speeds up the evolution of the system and decreases the tidal 
timescales.  Using up to order $\ell=6$ compared to $\ell=2$ results in the spin rates of the 
components changing up to 28\% faster at 2$R_{\rm p}$, but falling off quickly with increasing 
separation (Fig.~\ref{fig:wdot}) to less than 4\% at 5$R_{\rm p}$.  The size ratio of the components 
only affects $\dot{\omega}_{\rm s}$, where the higher-order terms are weaker for smaller secondaries.  
Similarly, for the change in semimajor axis with time, assuming both components are causing the 
separation to change in the same sense (${\rm sign}\left(\omega_{\rm p}-n\right)$ and 
${\rm sign}\left(\omega_{\rm s}-n\right)$ have the same value), using up to order $\ell=6$ 
(Fig.~\ref{fig:adot}) results in a faster evolution by 21\% to 28\% at 2$R_{\rm p}$ and decreases 
quickly with increasing separation.  Unlike the changes in spin rates, the largest effect on the 
evolution of the semimajor axis occurs when the size ratio is either unity (equal size) or negligible 
or when the spin of the secondary has synchronized to the mean motion such that the tidal torque 
on the secondary vanishes.  The change in semimajor axis with time is least affected by the 
higher-order terms for a size ratio of 0.53 with all other size ratios falling within these bounds.  
According to (\ref{eq:ndot}) for the change in the mean motion with time, the value of $\dot{n}/n$ 
using higher-order terms compared to $\ell=2$ has the same form as the change in semimajor axis in 
Fig.~\ref{fig:adot}.

The strengths of the contributions of the extra terms in the close-orbit correction to the change in 
semimajor axis are listed in Table~\ref{tab:astrength}.  At 2$R_{\rm p}$, higher-order terms with 
$\ell \ge 3$ account for nearly 25\% of the change in semimajor axis with time.  Although the $\ell=6$ 
term is necessary for accurate reproduction of the potential between the bodies to within 1\% at 
2$R_{\rm p}$, it does not alter the change in semimajor axis with time at the 1\% level because of the 
stronger dependence of (\ref{eq:adotp}) on separation compared to (\ref{eq:Vlegendre}).  The net 
contribution of the higher-order terms in Table~\ref{tab:astrength} decreases by roughly 5\% at each 
value of the separation from Table~\ref{tab:legendre} with only the $\ell=3$ term having much 
consequence beyond 3$R_{\rm p}$.  

\begin{figure}[!ht]
\begin{center}
\includegraphics[angle=0., scale=0.8]{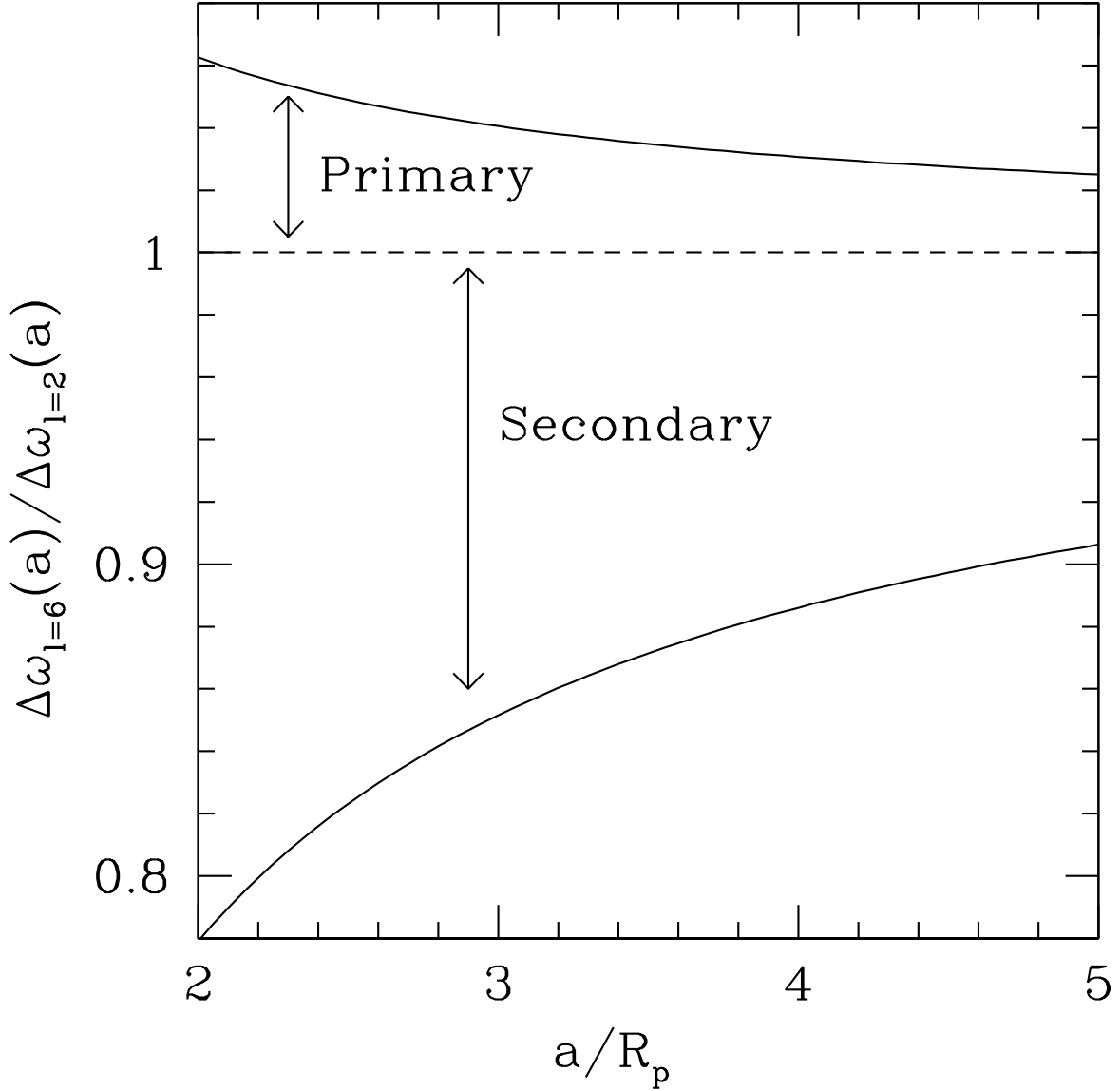}
\caption[]{Total change in spin rate of the components based on tidal evolution from an initial 
separation of 2$R_{\rm p}$ using all orders up to $\ell=6$ versus using classical $\ell=2$ tides.  
The coordinate on the $x$-axis is the final separation of the tidal evolution.  With both components
contributing in an additive sense in (\ref{eq:adotboth}), the spin rate of the primary is affected 
more rapidly than in the classical case, while the secondary is affected less rapidly.  The maximum 
change in the spin rate of the primary occurs for $R_{\rm s}/R_{\rm p}=0.53$ ($\textit{upper solid 
curve}$) and the minimum is the $\textit{dashed line}$ at 1 for $R_{\rm s}/R_{\rm p}=0,1$ or a 
synchronized secondary.  The $\textit{lower solid curve}$ corresponds to the change in spin rate of 
the secondary for $R_{\rm s}/R_{\rm p} \rightarrow 0$.  For larger size ratios, the curve for the 
secondary moves toward the $\textit{dashed line}$ at 1.}
\label{fig:wa}
\end{center}
\end{figure}

The total change in the component spin rates as a function of separation, shown in Fig.~\ref{fig:wa}, 
is given by integration of the ratio of~(\ref{eq:wpall}) and~(\ref{eq:adotboth}) for the primary 
and the ratio of~(\ref{eq:wsall}) and~(\ref{eq:adotboth}) for the secondary.  Depending on the size 
ratio of the components, the total change in the spin rate of the primary is enhanced by up to 6\% at 
2$R_{\rm p}$ over using $\ell=2$ tides only, but not by more than a few percent at larger separations.  
For the secondary, perhaps counter-intuitively, despite the spin of the secondary evolving more rapidly 
with time by adding higher-order terms (Fig.~\ref{fig:wdot}), its evolution with respect to the 
separation is less than when using $\ell=2$ only; the deficit is as large as 22\% at 2$R_{\rm p}$ when 
the size of the secondary is negligible.  This is because for smaller secondaries, the effect of 
higher-order terms on $\dot{\omega}_{\rm s}$ in~(\ref{eq:wsall}) is reduced, while the effect of 
higher-order terms on $\dot{\omega}_{\rm p}$ is independent of the size ratio.  Thus, for a rapidly 
rotating primary, the higher-order terms transfer more angular momentum from the spin of the primary 
to the orbit, expanding the separation faster than by $\ell=2$ tides alone and faster than the spin 
rate of the secondary changes such that the net effect on $\Delta\omega_{\rm s}(a)$ is smaller.

\begin{figure}[!ht]
\begin{center}
\includegraphics[angle=0., scale=0.8]{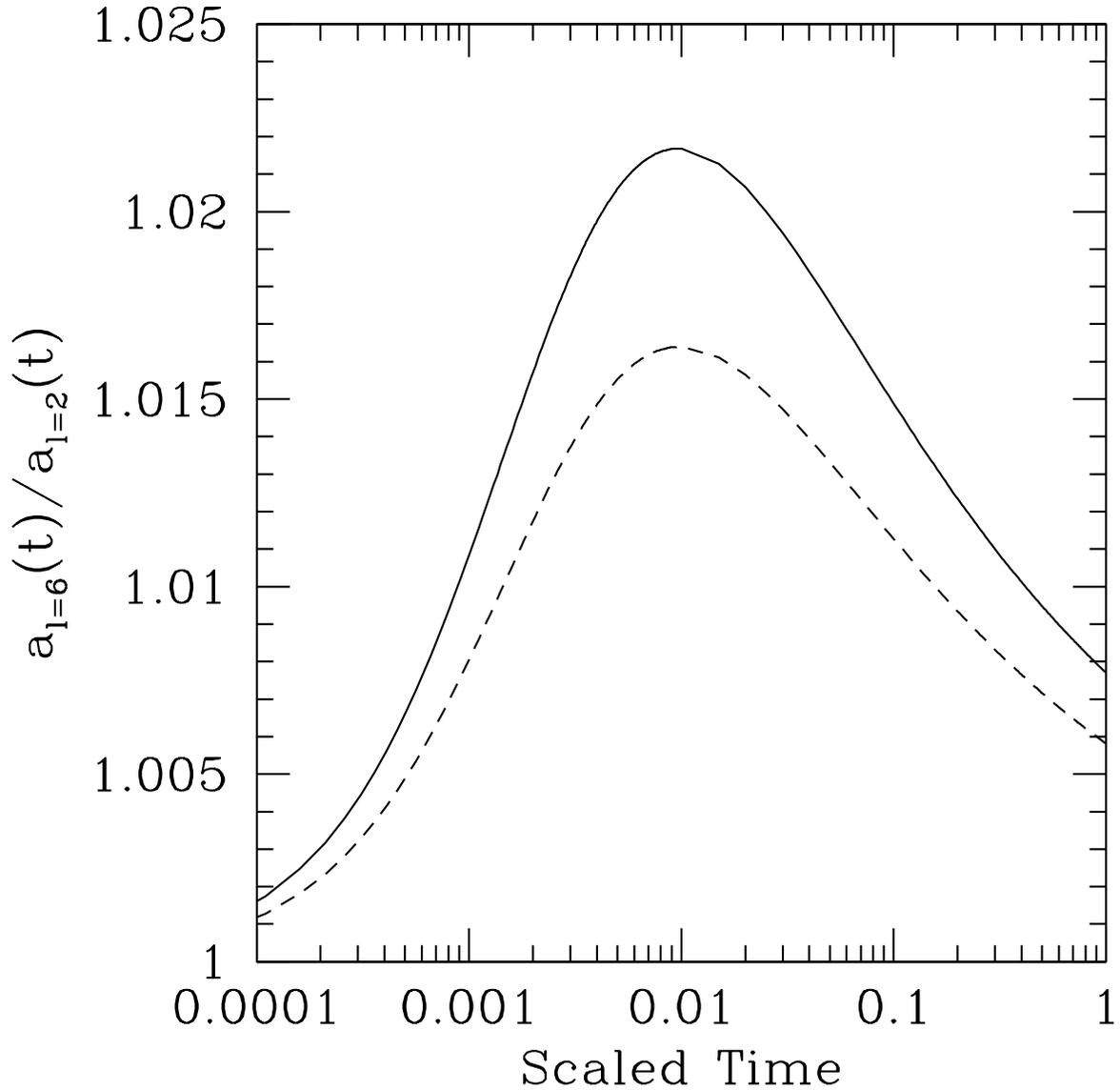}
\caption[]{Evolution of the semimajor axis with time using all orders up to $\ell=6$ versus using
classical $\ell=2$ tides.  Time is plotted logarithmically and scaled to the time necessary for a 
system to evolve from 2$R_{\rm p}$ to 5$R_{\rm p}$ via $\ell=2$ tides.  As in Fig.~\ref{fig:adot}, 
the $\textit{solid curve}$ corresponds to $R_{\rm s}/R_{\rm p}=0,1$ or a synchronized secondary and 
the $\textit{dashed curve}$ corresponds to $R_{\rm s}/R_{\rm p}=0.53$.  Using up to $\ell=6$ gives 
a correction of order 1\% to classical tides at any point in the evolution from 2$R_{\rm p}$ to 
5$R_{\rm p}$.}
\label{fig:at}
\end{center}
\end{figure}

Integration of~(\ref{eq:adotboth}) provides the separation as a function of time.  For tidal 
evolution from an initial separation of 2$R_{\rm p}$ to a final separation of 5$R_{\rm p}$ 
(Fig.~\ref{fig:at}), the close-orbit correction is strongest at the onset, expanding the 
separation more rapidly than $\ell=2$ tides, but only by about 2\% over the same time interval.  
The contributions from the higher-order terms lose strength over time as the separation increases 
resulting in a net effect of expanding the separation by $\sim$1\% extra by using $\ell=6$ 
instead of $\ell=2$.  From Figs.~\ref{fig:wa} and~\ref{fig:at}, the integrated effects of the 
close-orbit correction are small, typically of order a few percent; the effects are more 
noticeable in the instantaneous rates of change of the spin rates, separation, or mean motion 
due to the rapid fall-off in strength of the higher-order terms with increasing separation and 
how rapidly the system tidally evolves from small separations.  

Rearrangement and integration of (\ref{eq:adotboth}) allows one to calculate the combination of the 
material properties of the components $\mu Q$ (assuming $\mu_{\rm p}Q_{\rm p}=\mu_{\rm s}Q_{\rm s}$) 
and the age of the binary $\Delta t$ based on measurable system parameters.  For brevity, we retain 
only terms due to tides raised on the primary giving

\begin{eqnarray}
\frac{\mu Q}{\Delta t} & = & \frac{8\sqrt{3}}{19}\pi^{3/2}G^{3/2}\rho_{\rm p}^{5/2}R_{\rm p}^{2}\,\kappa\left(1+\kappa\right)^{1/2}\nonumber\\
 & & \times \left[\int_{2}^{a_{\rm f}/R_{\rm p}}\frac{x^{11/2}}{1+\frac{19}{22}x^{-2}+\frac{380}{459}x^{-4}+\frac{475}{584}x^{-6}+\frac{133}{165}x^{-8}}\,{\rm d}x\right]^{-1}
\label{eq:muq}
\end{eqnarray}

\noindent
with $x=a/R_{\rm p}$.  Because both terms on the left-hand side of (\ref{eq:muq}) are unknown, one 
may either estimate the material properties by assuming binary ages~\citep{marg02s,marg03,tayl07dpstides},
estimate binary ages by assuming material properties~\citep{wals06,gold09}, or consider both
avenues~\citep{marc08ecc,marc08circ,tayl10material}.  Furthermore, precisely because both terms are 
unknown, assuming a value for one has an intimate effect on the calculation of the other as changing 
one's value by an order of magnitude changes the result of the other by an order of magnitude.  Thus, 
when one wishes to find $\mu Q$, for instance, choosing an age for the binary injects a great source 
of uncertainty into the calculation.  

\begin{figure}[!ht]
\begin{center}
\includegraphics[angle=0., scale=0.8]{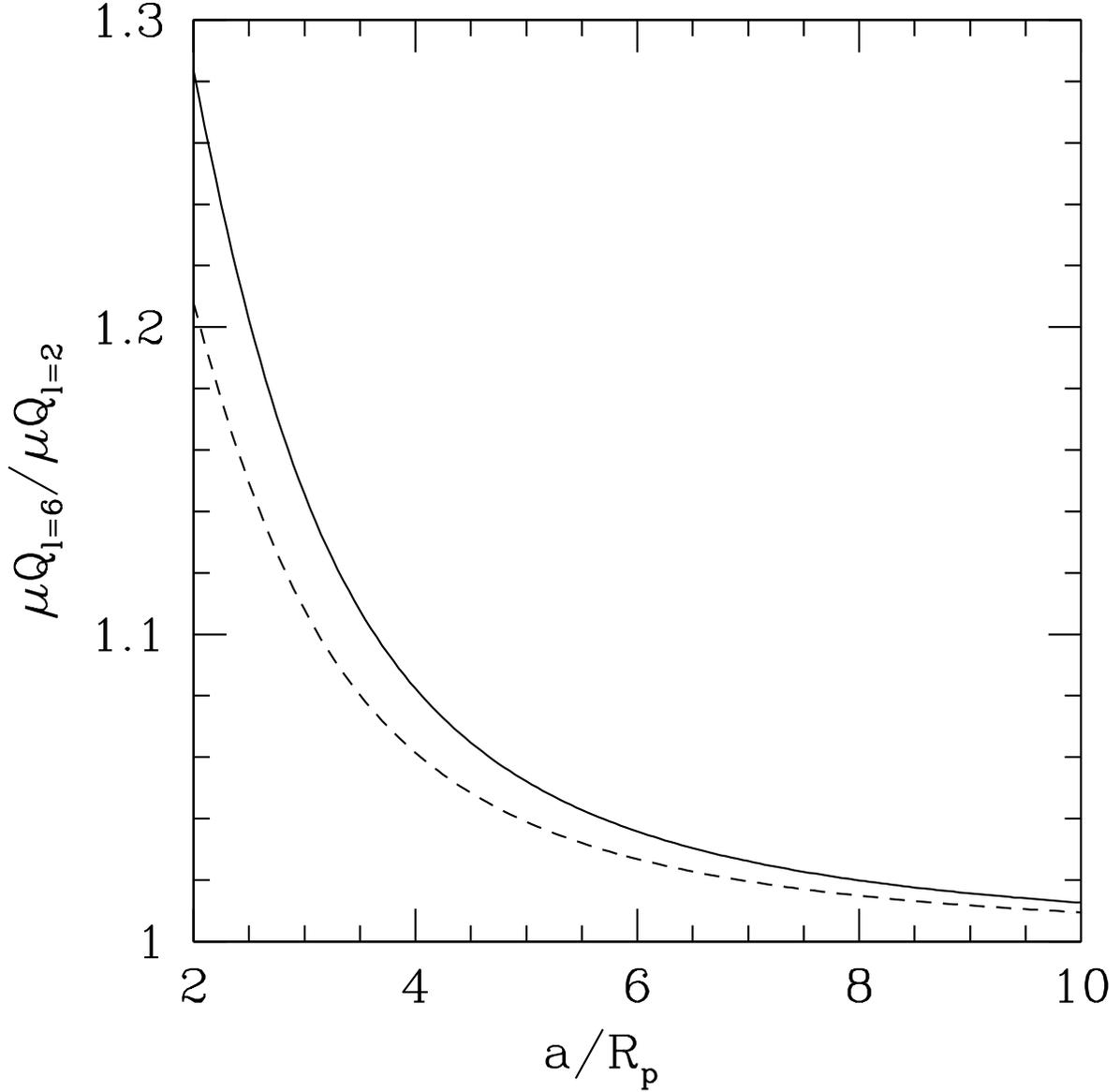}
\caption[]{In terms of the ratio $\mu Q_{\ell=6}/\mu Q_{\ell=2}$ for tidal evolution from 2$R_{\rm p}$ with 
both components contributing, systems with $R_{\rm s}/R_{\rm p}=0,1$ or a synchronized secondary share the 
$\textit{upper curve}$ and are most affected by the close-orbit correction; a system with 
$R_{\rm s}/R_{\rm p}=0.53$ ($\textit{dashed curve}$) is the least affected by the close-orbit correction.  
Overall, the close-orbit correction is roughly 25\% at 2$R_{\rm p}$ and quickly falls off to 5\% for 
evolution from 2$R_{\rm p}$ to 5$R_{\rm p}$ and to 1\% for evolution to 10$R_{\rm p}$ for all size ratios.  
The components are assumed to have similar $\mu Q$ parameters and contribute in an additive sense in 
(\ref{eq:adotboth}).}
\label{fig:muq}
\end{center}
\end{figure}

The close-orbit correction enhances the rate at which the separation changes such that, to provide 
the same tidal evolution over the same timescale $\Delta t$, the product $\mu Q$ must increase to 
compensate for the inclusion of the higher-order terms.  For classical $\ell=2$ tides, the denominator 
of the integrand in (\ref{eq:muq}) vanishes such that the effect of including terms up to $\ell=6$ 
alters $\mu Q$ according to

\begin{equation}
\frac{\mu Q_{\ell=6}}{\mu Q_{\ell=2}}~=~\frac{\int_{2}^{a_{\rm f}/R_{\rm p}}x^{11/2}\,{\rm d}x}{\int_{2}^{a_{\rm f}/R_{\rm p}}\frac{x^{11/2}}{1+\frac{19}{22}x^{-2}+\frac{380}{459}x^{-4}+\frac{475}{584}x^{-6}+\frac{133}{165}x^{-8}}\,{\rm d}x}
\label{eq:muqchange}
\end{equation} 

\noindent
and is shown as a function of the final separation in Fig.~\ref{fig:muq}.  Note that in 
Fig.~\ref{fig:muq}, the contribution of the secondary is included in the numerical integration of 
(\ref{eq:adotboth}) although it is not explicitly given in (\ref{eq:muqchange}) above.  Evolution 
from a close initial separation of 2$R_{\rm p}$ to a wide separation of 10$R_{\rm p}$ results in 
only a $\sim$1\% increase in $\mu Q$ over the classical value for all size ratios.  Thus, the basic 
$\ell=2$ tidal mechanism is sufficient for well-separated binaries.  On the other hand, if the final 
separation is smaller, as is the case for most near-Earth binaries, the correction is larger, 
increasing to 5\% for evolution from 2$R_{\rm p}$ to 5$R_{\rm p}$ and 15\% for evolution from 
2$R_{\rm p}$ to 3$R_{\rm p}$.  When making a coarse estimate of the material properties of the 
system, taking the close orbit into account is not of paramount importance; classical tides will 
easily provide an order-of-magnitude estimate of $\mu Q$ for even the closest of binary asteroids, 
though the result will be slightly underestimated.  Complementarily, if higher-order terms are 
included and $\mu Q$ is held fixed, the age of the binary must decrease by the same factor as in 
Fig.~\ref{fig:muq} meaning that $\ell=2$ tides provide an upper bound on ages for systems with a 
given $\mu Q$ value.  

The use of higher-order terms up to $\ell=6$ is sufficient for exploring the tidal evolution of 
binary systems with separations greater than 2$R_{\rm p}$.  Additional terms with $\ell>6$ make
inconsequential changes to tidal evolution at these separations as illustrated by the rapid fall-off 
of the contributions of the higher-order terms beyond 2$R_{\rm p}$ in Table~\ref{tab:astrength}.  
Moreover, terms with $\ell>6$ leave Figs.~\ref{fig:wdot}--\ref{fig:muq} unchanged, only having an 
effect within 2$R_{\rm p}$.  Thus, if one wishes to proceed inward of 2$R_{\rm p}$, simply using 
orders of up to $\ell=6$ is insufficient as higher-order terms gain importance the closer one 
proceeds to the primary.  Though we stated earlier that the number of terms required can rapidly 
become unwieldy, one can approximate their strength.  For an arbitrary order $\ell > 2$, the 
term within the square brackets of (\ref{eq:adotboth}) is approximately

\begin{equation}
0.8\left(\frac{a}{R_{\rm p}}\right)^{-2\left(\ell-2\right)}\left({\rm sign}\left(\omega_{\rm p}-n\right)+\left(\frac{R_{\rm s}}{R_{\rm p}}\right)^{2\ell-3}\frac{\mu_{\rm p} Q_{\rm p}}{\mu_{\rm s} Q_{\rm s}}{\rm sign}\left(\omega_{\rm s}-n\right)\right),
\label{eq:xterm}
\end{equation}

\noindent
allowing additional terms to be included without explicit calculation of the Love numbers 
$k_{\ell, \rm p}$ or manipulation of the Legendre polynomials.  Similar terms follow for the 
changes in spin rates.  One must keep in mind that the approximation in~(\ref{eq:xterm}) is only 
valid so long as the small angle approximation holds 
$\left(\sin\,m\delta\simeq m\delta\propto Q^{-1}\right)$ with $m\le\ell$, which requires $Q>10$ to 
retain 1\% accuracy at $m=6$ and larger $Q$ as $m$ increases\footnote{We have applied (\ref{eq:Q}) to
estimate the value of $Q$ required.  In general, the small angle approximation holds to within 1\% for
$Q_{\ell m p q} \sim 4$ or greater.} (e.g., $Q>20$ for $m=10$).  Also, having separations of 
less than 2$R_{\rm p}$ requires smaller secondaries, since contact occurs at a separation of 
$(1+R_{\rm s}/R_{\rm p})\,R_{\rm p}$, which reduces the contribution of the secondary due to dependencies 
upon the size ratio, in addition to demanding consideration of the Roche limit for the system (see 
Section~\ref{sec:roche}).


\section{Comparison to Measurement Errors}

Take, for example, 66391 (1999 KW4), the best-studied of the near-Earth binary 
systems~\citep{ostr06,sche06s}.  Even with exhaustive analysis of radar imagery, production of 
three-dimensional shape models of both components, and investigation of the system dynamics, physical
parameters of the system are not known with extreme precision.  The densities of the primary and 
secondary components are known to approximately 12\% and 25\%, respectively.  The uncertainty in the
density of the primary alone can cause error of more than 30\% in $\dot{\omega}_{\rm p}$, 
$\dot{\omega}_{\rm s}$, and $\dot{a}/R_{\rm p}$ according to (\ref{eq:wpall}--\ref{eq:adotboth}), more 
than the close-orbit correction causes in Figs.~\ref{fig:wdot} and~\ref{fig:adot}.  The higher 
estimated density of the secondary in the 1999 KW4 system of 2.81 g/cm$^{3}$, compared to 1.97 g/cm$^{3}$
for the primary, directly affects the mass ratio $\kappa$ applied in the equations of tidal evolution as 
one typically assumes similar densities for the components.  Ignoring the density uncertainties, this 
difference in component densities alone causes a 43\% change in $\kappa$ that, in turn, affects 
$\dot{\omega}_{\rm p}$ by a factor of two and $\dot{\omega}_{\rm s}$ and $\dot{a}/R_{\rm p}$ by 
approximately 40\% as well, again, a larger effect than the close-orbit correction to tidal evolution.  
The calculated value of $\mu Q$ in~(\ref{eq:muq}) is affected by density and mass ratio uncertainties 
in the same way as $\dot{a}/R_{\rm p}$.  Furthermore, uncertainties in densities and the dependence 
of the mass ratio $\kappa$ on density differences between the components apply at all separations 
unlike the close-orbit correction, which falls off quickly with increasing separation.   

One must also consider the effect of the initial separation of the components at the onset of tidal
evolution, a property that is not known for individual systems, but can be estimated from simulations 
of binary formation mechanisms [e.g.,~\citet{wals06,wals08yorp}] and given a lower bound by 
the contact limit at $\left(1+R_{\rm s}/R_{\rm p}\right)\,R_{\rm p}$.  Assuming evolution over the 
same timescale, if the system had an initial separation $a_{\rm i}$ instead of 2$R_{\rm p}$, the 
effect on $\mu Q$ calculated with classical $\ell=2$ tides raised only on the primary is

\begin{equation}
\frac{\mu Q_{\rm i}}{\mu Q_{2}}~=~\frac{1-\left(\frac{2}{a_{\rm f}/R_{\rm p}}\right)^{13/2}}{1-\left(a_{\rm i}/a_{\rm f}\right)^{13/2}}.
\label{eq:initsep}
\end{equation}

\noindent
For a final separation $a_{\rm f}$ from 3$R_{\rm p}$ to 10$R_{\rm p}$, unless the actual initial 
separation $a_{\rm i}$ is within 10\% of the final separation ($>$0.9$a_{\rm f}$), the value of $\mu Q$ 
is affected by less than a factor of two by assuming an initial separation of 2$R_{\rm p}$.  Using up 
to $\ell=6$ and allowing for tides raised on the secondary with any size ratio do not cause a significant 
difference in this result. 

A similar result is found for the dependence on the final separation of the components, which one 
typically takes to be the current separation.  If $a_{\rm f}$ is the final (current) separation, 
then changing the separation to $a_{\rm f^{\,\prime}}$ due to, say, a measurement error causes the 
calculated $\mu Q$ value for tidal evolution from 2$R_{\rm p}$ to change as 

\begin{equation}
\frac{\mu Q_{\rm f^{\,\prime}}}{\mu Q_{\rm f}}~=~\frac{1-\left(\frac{2}{a_{\rm f}/R_{\rm p}}\right)^{13/2}}{\left(a_{\rm f^{\,\prime}}/a_{\rm f}\right)^{13/2}-\left(\frac{2}{a_{\rm f}/R_{\rm p}}\right)^{13/2}}.
\label{eq:finalsep}
\end{equation}

\noindent
We find $\mu Q$ is affected by less than a factor of two if the final (current) separation is known 
within 10\%.  From the dependence on the initial and final separations, it is clear that the tidal 
evolution near the final separation dominates over the early evolution where the close-orbit correction 
is necessary.  In fact, if instead of calculating $\mu Q$, one considers the time taken to tidally 
evolve to a final separation $a_{\rm f} \ge 4R_{\rm p}$ (by assuming a value of $\mu Q$ instead of an age), 
the evolution of the separation from 0.9$a_{\rm f}$ to $a_{\rm f}$ takes roughly the same amount of time 
as the evolution from $a_{\rm i}\le 2R_{\rm p}$ to 0.9$a_{\rm f}$.  Thus, precisely when the close-orbit 
correction is most prominent is also when the system requires the least amount of time to evolve, which 
causes the mild effect of the close-orbit correction found in Figs.~\ref{fig:at} and~\ref{fig:muq}.

Returning to a concrete example, for the 1999 KW4 system, using the equivalent spherical radius of 
the primary shape model, the separation of the components $a/R_{\rm p}$ is known to 3\% as 
3.87$\pm$0.12~\citep{ostr06}.  By~(\ref{eq:finalsep}), this small uncertainty can result in a roughly 
20\% error in the calculated $\mu Q$, more than twice the effect of the close-orbit correction in 
Fig.~\ref{fig:muq} at 3.87$R_{\rm p}$.  Together with the dependence of the $\mu Q$ calculation on 
the density values for the components, the accuracy of measurements of physical parameters in the 
1999 KW4 system is more important than accounting for the proximity of the components to one another.


\section{Discussion}
\label{sec:disc}

We have derived the equations of tidal evolution to arbitrary order in the Legendre polynomial expansion
of the separation between two spherical bodies in a circular and equatorial mutual orbit allowing for 
accurate representation of evolution within five primary radii.  Equations written in terms of the Love 
number $k_{\ell}$ are applicable to any binary system, while equations where the Love number has been 
evaluated have assumed the bodies involved have rigidities that dominate their self-gravitational stress 
(characteristic of bodies less than roughly 200~km in radius).  Because higher-order terms cause tidal 
evolution to proceed faster, choosing to ignore them produces upper limits on tidal evolution timescales 
and lower limits on material properties in terms of the product of rigidity and the tidal dissipation 
function.  However, we have shown that the correction for close orbits has only a minor integrated effect 
on outward tidal evolution and the calculation of material properties, comparable to or less than the 
effect of uncertainties in measurable properties such as density, mass ratio, and semimajor axis (scaled 
to the radius of the primary component).  In the case of outward evolution, the binary system evolves 
rapidly through the range of separations where the close-orbit correction is strongest, so one can safely 
ignore the correction to obtain order-of-magnitude estimates of timescales and material properties using 
the classical equations for tidal evolution.  Accounting for higher orders is more applicable to studying, 
famously in the case of Phobos, observed secular accelerations and the infall of a secondary to the surface 
of its primary where the higher-order terms instead gain strength.  

Though we have presented the expansion of the gravitational potential and the resulting equations of tidal
evolution in the context of two asteroids in mutual orbit, the essence of this work could be generalized 
for use in the determination of the Roche limit and the study of close flybys.  The use of a higher-order 
expansion of the gravitational potential in terms of Legendre polynomials is warranted whenever the 
separation of two bodies is within five times the radius of one of the bodies\footnote{The potential felt 
by the primary requires higher-order terms with $\ell>2$ if the separation is less than  5$R_{\rm p}$; 
the potential felt by the secondary requires higher-order terms with $\ell>2$ if the separation is less than 
5$R_{\rm s}$.} (see Table~\ref{tab:legendre}).  Historically, in the context of disruption of a body at the 
Roche limit or due to a close flyby of a larger body~\citep{srid92,rich98,hols06,hols08,shar06,shar09}, 
stresses are only considered in the much smaller secondary while the primary is assumed to be rigid.  For 
small secondaries, the cohesionless Roche limit of 1.5--2$R_{\rm p}$ is much larger than 5$R_{\rm s}$ such 
that higher-order terms in the potential expansion are not necessary.  However, as larger secondaries are 
considered ($R_{\rm s}/R_{\rm p}>0.1$), higher-order terms in the gravitational potential will further stress 
the secondary near the Roche limit.  Also, with components of increasingly similar size, the assumption of a 
rigid primary is not appropriate; the tidal stress on the primary will deform it from a spherical shape and 
produce an external potential as in Section~\ref{sec:extpotl} that will in turn further stress the secondary.  
If the components are not spin-locked, tidal torques will also play a role in stressing the secondary.  Thus, 
if evaluating the Roche limit for components of similar size and/or components that are not spin-locked, 
one must consider the description presented here.  For disruption during a close flyby, or simply modification 
of the spin state of the passing body~\citep{sche01,sche00,sche04}, one must consider the proximity of the 
flyby in terms of the expansion of the gravitational potential and whether or not tidal bulges can be raised 
on the components that would produce torques capable of further altering the spin state of either component.

It is also important to remember that the higher-order theory presented here has implicitly assumed initially 
spherical bodies.  Extension of this work from spheres to ellipsoids or to arbitrary shapes would affect the 
mutual gravitational potential, linear and angular momentum balance, and orbital equations as described
by~\citet{sche09} and~\citet{shar10}.  Once the shape is made nonspherical in the absence of a tidal 
potential, the system is subject to a ``direct'' torque that naturally occurs from the changing gravitational 
pull felt by the orbiting component due to the nonspherical shape of the other component.  Accounting for the 
tidal potential introduces the ``indirect'' torque described here due to the deformation of one component 
by the gravitational presence of the other component.  Because the amplitude of the tidal bulge on asteroids, 
the parameter $\lambda$ in this work, can be very small due to its direct dependence on the ratio of 
self-gravitational stress to rigidity, its direct dependence on the mass ratio, and its inverse dependence 
on the separation raised to the third (or higher) power, natural deviations from a spherical shape may exceed 
the amplitude of the tidal bulge.  However, one must recall that the direct torques due to a nonspherical 
shape will change direction as the body rotates under the orbiting component tending to cancel the pre- and 
post-encounter effects of the torque as opposed to the indirect torque that is in a consistent direction so 
long as the bulge always leads or lags the orbiting component.  It may be important to consider direct torques 
due to natural departures from a spherical shape via the use of shape models:  oblate or prolate spheroids, 
triaxial ellipsoids, or vertex models such as those made for the components of the 1999 KW4 binary system and 
other asteroids.


\section*{Acknowledgements}

The authors are indebted to the two referees whose detailed reviews and insightful suggestions improved the 
clarity and quality of the manuscript.  The authors are especially grateful to Michael Efroimsky for many 
discussions on the finer points of tidal theory and celestial mechanics.  This work was supported by NASA 
Planetary Astronomy grants NNG04GN31G and NNX07AK68G to Jean-Luc Margot.


\bibliographystyle{icarus}
\bibliography{TaylorMargot-CloseTides}

\begin{thebibliography}{}

\bibitem[{A'Hearn} et~al.(2005){A'Hearn}, {Belton}, {Delamere}, {Kissel},
  {Klaasen}, {McFadden}, {Meech}, {Melosh}, {Schultz}, {Sunshine}, {Thomas},
  {Veverka}, {Yeomans}, {Baca}, {Busko}, {Crockett}, {Collins}, {Desnoyer},
  {Eberhardy}, {Ernst}, {Farnham}, {Feaga}, {Groussin}, {Hampton}, {Ipatov},
  {Li}, {Lindler}, {Lisse}, {Mastrodemos}, {Owen}, {Richardson}, {Wellnitz},
  and {White}]{aher05}
{A'Hearn}, M.~F., and 32 colleagues, 2005.
\newblock {Deep Impact: Excavating comet Tempel 1}.
\newblock Science~310, 258--264.

\bibitem[{Behrend} et~al.(2006){Behrend}, {Bernasconi}, {Roy}, {Klotz},
  {Colas}, {Antonini}, {Aoun}, {Augustesen}, {Barbotin}, {Berger},
  {Berrouachdi}, {Brochard}, {Cazenave}, {Cavadore}, {Coloma}, {Cotrez},
  {Deconihout}, {Demeautis}, {Dorseuil}, {Dubos}, {Durkee}, {Frappa},
  {Hormuth}, {Itkonen}, {Jacques}, {Kurtze}, {Laffont}, {Lavayssi{\`e}re},
  {Lecacheux}, {Leroy}, {Manzini}, {Masi}, {Matter}, {Michelsen}, {Nomen},
  {Oksanen}, {P{\"a}{\"a}kk{\"o}nen}, {Peyrot}, {Pimentel}, {Pray}, {Rinner},
  {Sanchez}, {Sonnenberg}, {Sposetti}, {Starkey}, {Stoss}, {Teng}, {Vignand},
  and {Waelchli}]{behr06}
{Behrend}, R., and 48 colleagues, 2006.
\newblock {Four new binary minor planets: (854) Frostia, (1089) Tama, (1313)
  Berna, (4492) Debussy}.
\newblock Astron. Astroph.~446, 1177--1184.

\bibitem[{Bills} et~al.(2005){Bills}, {Neumann}, {Smith}, and {Zuber}]{bill05}
{Bills}, B.~G., {Neumann}, G.~A., {Smith}, D.~E., {Zuber}, M.~T., 2005.
\newblock {Improved estimate of tidal dissipation within Mars from MOLA
  observations of the shadow of Phobos}.
\newblock J. Geophys. Res.~110, 2376--2406.

\bibitem[{Burns}(1976){Burns}]{burn76}
{Burns}, J.~A., 1976.
\newblock {Elementary derivation of the perturbation equations of celestial
  mechanics}.
\newblock Am. J. Phys.~44, 944--949.

\bibitem[{Burns}(1977){Burns}]{burn77}
{Burns}, J.~A., 1977.
\newblock {Orbital evolution}.
\newblock In: Burns, J.~A. (Ed.), Planetary Satellites. Univ. of Arizona Press,
  Tucson, pp.\  113--156.

\bibitem[{Chandrasekhar}(1969){Chandrasekhar}]{chan69}
{Chandrasekhar}, S., 1969.
\newblock {\em {Ellipsoidal Figures of Equilibrium}}.
\newblock Yale Univ. Press, New Haven.

\bibitem[{Danby}(1992){Danby}]{danb92}
{Danby}, J.~M.~A., 1992.
\newblock {\em {Fundamentals of Celestial Mechanics}}.
\newblock Willman-Bell, Richmond.

\bibitem[{Darwin}(1879a){Darwin}]{darw79a}
{Darwin}, G.~H., 1879a.
\newblock {On the bodily tides of viscous and semi-elastic spheroids, and on
  the ocean tides upon a yielding nucleus}.
\newblock Philos. Trans. R. Soc. London~170, 1--35.

\bibitem[{Darwin}(1879b){Darwin}]{darw79b}
{Darwin}, G.~H., 1879b.
\newblock {On the precession of a viscous spheroid, and on the remote history
  of the Earth}.
\newblock Philos. Trans. R. Soc. London~170, 447--538.

\bibitem[{Darwin}(1880){Darwin}]{darw80}
{Darwin}, G.~H., 1880.
\newblock {On the secular changes in the elements of the orbit of a satellite
  revolving about a tidally distorted planet}.
\newblock Philos. Trans. R. Soc. London~171, 713--891.

\bibitem[{Descamps} and {Marchis}(2008){Descamps} and {Marchis}]{desc08}
{Descamps}, P., {Marchis}, F., 2008.
\newblock {Angular momentum of binary asteroids: Implications for their
  possible origin}.
\newblock Icarus~193, 74--84.

\bibitem[{Descamps} et~al.(2007){Descamps}, {Marchis}, {Michalowski},
  {Vachier}, {Colas}, {Berthier}, {Assafin}, {Dunckel}, {Polinska}, {Pych},
  {Hestroffer}, {Miller}, {Vieira-Martins}, {Birlan}, {Teng-Chuen-Yu},
  {Peyrot}, {Payet}, {Dorseuil}, {L{\'e}onie}, and {Dijoux}]{desc07}
{Descamps}, P., and 19 colleagues, 2007.
\newblock {Figure of the double asteroid 90 Antiope from adaptive optics and
  lightcurve observations}.
\newblock Icarus~187, 482--499.

\bibitem[{Efroimsky} and {Williams}(2009){Efroimsky} and {Williams}]{efro09}
{Efroimsky}, M., {Williams}, J.~G., 2009.
\newblock {Tidal torques: A critical review of some techniques}.
\newblock Cel. Mech. Dyn. Astron.~104, 257--289.

\bibitem[{Ferraz-Mello} et~al.(2008){Ferraz-Mello}, {Rodr{\'{\i}}guez}, and
  {Hussmann}]{ferr08}
{Ferraz-Mello}, S., {Rodr{\'{\i}}guez}, A., {Hussmann}, H., 2008.
\newblock {Tidal friction in close-in satellites and exoplanets: The Darwin
  theory re-visited}.
\newblock Celest. Mech. Dyn. Astron.~101, 171--201.

\bibitem[{Gerstenkorn}(1955){Gerstenkorn}]{gers55}
{Gerstenkorn}, H., 1955.
\newblock {{\"U}ber Gezeitenreibung beim Zweik{\"o}rperproblem}.
\newblock Zeitschrift fur Astrophysik~36, 245--274.

\bibitem[{Goldreich}(1963){Goldreich}]{gold63}
{Goldreich}, P., 1963.
\newblock {On the eccentricity of satellite orbits in the solar system}.
\newblock Mon. Not. R. Astron. Soc.~126, 257--268.

\bibitem[{Goldreich}(1966){Goldreich}]{gold66moon}
{Goldreich}, P., 1966.
\newblock {History of the lunar orbit}.
\newblock Rev. Geophys. Space Phys.~4, 411--439.

\bibitem[{Goldreich} and {Soter}(1966){Goldreich} and {Soter}]{gold66}
{Goldreich}, P., {Soter}, S., 1966.
\newblock {Q in the solar system}.
\newblock Icarus~5, 375--389.

\bibitem[{Goldreich} and {Sari}(2009){Goldreich} and {Sari}]{gold09}
{Goldreich}, P., {Sari}, R., 2009.
\newblock {Tidal evolution of rubble piles}.
\newblock Astroph. J.~691, 54--60.

\bibitem[{Holsapple} and {Michel}(2006){Holsapple} and {Michel}]{hols06}
{Holsapple}, K.~A., {Michel}, P., 2006.
\newblock {Tidal disruptions: A continuum theory for solid bodies}.
\newblock Icarus~183, 331--348.

\bibitem[{Holsapple} and {Michel}(2008){Holsapple} and {Michel}]{hols08}
{Holsapple}, K.~A., {Michel}, P., 2008.
\newblock {Tidal disruptions. II. A continuum theory for solid bodies with
  strength, with applications to the Solar System}.
\newblock Icarus~193, 283--301.

\bibitem[{Kaula}(1964){Kaula}]{kaul64}
{Kaula}, W.~M., 1964.
\newblock Tidal dissipation by solid friction and the resulting orbital
  evolution.
\newblock Rev. Geophys.~2, 661--685.

\bibitem[{Lambeck}(1979){Lambeck}]{lamb79}
{Lambeck}, K., 1979.
\newblock {On the orbital evolution of the Martian satellites}.
\newblock J. Geophys. Res.~84, 5651--5658.

\bibitem[{Love}(1927){Love}]{love27}
{Love}, A.~E.~H., 1927.
\newblock {\em {A Treatise on the Mathematical Theory of Elasticity}}.
\newblock Dover, New York.

\bibitem[{MacDonald}(1964){MacDonald}]{macd64}
{MacDonald}, G.~J.~F., 1964.
\newblock {Tidal friction}.
\newblock Rev. Geophys. Space Phys.~2, 467--541.

\bibitem[{MacRobert}(1967){MacRobert}]{macr67}
{MacRobert}, T.~M., 1967.
\newblock {\em {Spherical Harmonics}}.
\newblock Pergamon Press, Oxford.

\bibitem[{Marchis} et~al.(2008{\natexlab{a}})){Marchis}, {Descamps}, {Berthier}, {Hestroffer},
  {Vachier}, {Baek}, {Harris}, and {Nesvorn{\'y}}]{marc08ecc}
{Marchis}, F., {Descamps}, P., {Berthier}, J., {Hestroffer}, D., {Vachier}, F.,
  {Baek}, M., {Harris}, A.~W., {Nesvorn{\'y}}, D., 2008{\natexlab{a}}.
\newblock {Main belt binary asteroidal systems with eccentric mutual orbits}.
\newblock Icarus~195, 295--316.

\bibitem[{Marchis} et~al.(2008{\natexlab{b}}){Marchis}, {Descamps}, {Baek}, {Harris},
  {Kaasalainen}, {Berthier}, {Hestroffer}, and {Vachier}]{marc08circ}
{Marchis}, F., {Descamps}, P., {Baek}, M., {Harris}, A.~W., {Kaasalainen}, M.,
  {Berthier}, J., {Hestroffer}, D., {Vachier}, F., 2008{\natexlab{b}}.
\newblock {Main belt binary asteroidal systems with circular mutual orbits}.
\newblock Icarus~196, 97--118.

\bibitem[{Margot} et~al.(2002){Margot}, {Nolan}, {Benner}, {Ostro}, {Jurgens},
  {Giorgini}, {Slade}, and {Campbell}]{marg02s}
{Margot}, J.~L., {Nolan}, M.~C., {Benner}, L.~A.~M., {Ostro}, S.~J., {Jurgens},
  R.~F., {Giorgini}, J.~D., {Slade}, M.~A., {Campbell}, D.~B., 2002.
\newblock Binary asteroids in the near-Earth object population.
\newblock Science~296, 1445--1448.

\bibitem[{Margot} et~al.(2003){Margot}, {Nolan}, {Negron}, {Hine}, {Campbell},
  {Howell}, {Benner}, {Ostro}, {Giorgini}, and {Marsden}]{marg03}
{Margot}, J.~L., {Nolan}, M.~C., {Negron}, V., {Hine}, A.~A., {Campbell},
  D.~B., {Howell}, E.~S., {Benner}, L.~A.~M., {Ostro}, S.~J., {Giorgini},
  J.~D., {Marsden}, B.~G., 2003.
\newblock {1937 UB (Hermes)}.
\newblock IAU Circ.~8227, 2.

\bibitem[{Margot} et~al.(2006){Margot}, {Pravec}, {Nolan}, {Howell}, {Benner},
  {Giorgini}, F., {Ostro}, {Slade}, {Magri}, {Taylor}, {Nicholson}, and
  {Campbell}]{marg06iau}
{Margot}, J.~L., and 12 colleagues, 2006.
\newblock Hermes as an exceptional case among binary near-Earth asteroids.
\newblock In: IAU Gen. Assem.

\bibitem[{Merline} et~al.(2000){Merline}, {Close}, {Shelton}, {Dumas},
  {Menard}, {Chapman}, {Slater}, and {Keck II Telescope}]{merl00iauc}
{Merline}, W.~J., {Close}, L.~M., {Shelton}, J.~C., {Dumas}, C., {Menard}, F.,
  {Chapman}, C.~R., {Slater}, D.~C., {Keck II Telescope}, W.~M., 2000.
\newblock {Satellites of minor planets}.
\newblock IAU Circ.~7503, 3.

\bibitem[{Micha{\l}owski} et~al.(2004){Micha{\l}owski}, {Bartczak}, {Velichko},
  {Kryszczy{\'n}ska}, {Kwiatkowski}, {Breiter}, {Colas}, {Fauvaud},
  {Marciniak}, {Micha{\l}owski}, {Hirsch}, {Behrend}, {Bernasconi}, {Rinner},
  and {Charbonnel}]{mich04}
{Micha{\l}owski}, T., and 14 colleagues, 2004.
\newblock {Eclipsing binary asteroid 90 Antiope}.
\newblock Astron. Astroph.~423, 1159--1168.

\bibitem[{Mignard}(1979){Mignard}]{mign79}
{Mignard}, F., 1979.
\newblock {The evolution of the lunar orbit revisited. I}.
\newblock Moon and Planets~20, 301--315.

\bibitem[{Mignard}(1980){Mignard}]{mign80}
{Mignard}, F., 1980.
\newblock {The evolution of the lunar orbit revisited. II}.
\newblock Moon and Planets~23, 185--201.

\bibitem[{Mignard}(1981){Mignard}]{mign81}
{Mignard}, F., 1981.
\newblock {The lunar orbit revisited. III}.
\newblock Moon and Planets~24, 189--207.

\bibitem[{Munk} and {MacDonald}(1960){Munk} and {MacDonald}]{munk60}
{Munk}, W.~H., {MacDonald}, G.~J.~F., 1960.
\newblock {\em {The Rotation of the Earth}}.
\newblock Cambridge Univ. Press, Cambridge.

\bibitem[{Murray} and {Dermott}(1999){Murray} and {Dermott}]{murr99}
{Murray}, C.~D., {Dermott}, S.~F., 1999.
\newblock {\em Solar System Dynamics}.
\newblock Cambridge Univ. Press, Cambridge.

\bibitem[{Ostro} et~al.(2006){Ostro}, {Margot}, {Benner}, {Giorgini},
  {Scheeres}, {Fahnestock}, {Broschart}, {Bellerose}, {Nolan}, {Magri},
  {Pravec}, {Scheirich}, {Rose}, {Jurgens}, {De Jong}, and {Suzuki}]{ostr06}
{Ostro}, S.~J., and 15 colleagues, 2006.
\newblock {Radar imaging of binary near-Earth asteroid (66391) 1999 KW4}.
\newblock Science~314, 1276--1280.

\bibitem[{Peale}(1999){Peale}]{peal99}
{Peale}, S.~J., 1999.
\newblock {Origin and evolution of the natural satellites}.
\newblock Annu. Rev. Astron. Astrophys.~37, 533--602.

\bibitem[{Pravec} and {Harris}(2007){Pravec} and {Harris}]{prav07}
{Pravec}, P., {Harris}, A.~W., 2007.
\newblock {Binary asteroid population 1. Angular momentum content}.
\newblock Icarus~190, 250--259.

\bibitem[{Pravec} et~al.(2003){Pravec}, {Kusnirak}, {Warner}, {Behrend},
  {Harris}, {Oksanen}, {Higgins}, {Roy}, {Rinner}, {Demeautis}, {van den
  Abbeel}, {Klotz}, {Waelchli}, {Alderweireldt}, {Cotrez}, and
  {Brunetto}]{prav03}
{Pravec}, P., and 15 colleagues, 2003.
\newblock {1937 UB (Hermes)}.
\newblock IAU Circ.~8233, 3.

\bibitem[{Redmond} and {Fish}(1964){Redmond} and {Fish}]{redm64}
{Redmond}, J.~C., {Fish}, F.~F., 1964.
\newblock {The luni-tidal interval in Mars and the secular acceleration of
  Phobos}.
\newblock Icarus~3, 87--91.

\bibitem[{Richardson} and {Walsh}(2006){Richardson} and {Walsh}]{rich06}
{Richardson}, D.~C., {Walsh}, K.~J., 2006.
\newblock {Binary minor planets}.
\newblock Annu. Rev. Earth Planet. Sci.~34, 47--81.

\bibitem[{Richardson} et~al.(1998){Richardson}, {Bottke}, and {Love}]{rich98}
{Richardson}, D.~C., {Bottke}, W.~F., {Love}, S.~G., 1998.
\newblock {Tidal distortion and disruption of Earth-crossing asteroids}.
\newblock Icarus~134, 47--76.

\bibitem[{Richardson} et~al.(2007){Richardson}, {Melosh}, {Lisse}, and
  {Carcich}]{rich07}
{Richardson}, J.~E., {Melosh}, H.~J., {Lisse}, C.~M., {Carcich}, B., 2007.
\newblock {A ballistics analysis of the Deep Impact ejecta plume: Determining
  comet Tempel 1's gravity, mass, and density}.
\newblock Icarus~190, 357--390.

\bibitem[{Rubincam}(1973){Rubincam}]{rubi73}
{Rubincam}, D.~P., 1973.
\newblock {The early history of the lunar inclination}.
\newblock NASA-GSFC Rep. X-592-73-328, Goddard Space Flight Center, Greenbelt,
  Md.

\bibitem[{Rubincam}(2000){Rubincam}]{rubi00}
{Rubincam}, D.~P., 2000.
\newblock {Radiative spin-up and spin-down of small asteroids}.
\newblock Icarus~148, 2--11.

\bibitem[{Scheeres}(2001){Scheeres}]{sche01}
{Scheeres}, D.~J., 2001.
\newblock {Changes in rotational angular momentum due to gravitational
  interactions between two finite bodies}.
\newblock Cel. Mech. Dyn. Astron.~81, 39--44.

\bibitem[{Scheeres}(2009){Scheeres}]{sche09}
{Scheeres}, D.~J., 2009.
\newblock {Stability of the planar full 2-body problem}.
\newblock Cel. Mech. Dyn. Astron.~104, 103--128.

\bibitem[{Scheeres} et~al.(2000){Scheeres}, {Ostro}, {Werner}, {Asphaug}, and
  {Hudson}]{sche00}
{Scheeres}, D.~J., {Ostro}, S.~J., {Werner}, R.~A., {Asphaug}, E., {Hudson},
  R.~S., 2000.
\newblock {Effects of gravitational interactions on asteroid spin states}.
\newblock Icarus~147, 106--118.

\bibitem[{Scheeres} et~al.(2004){Scheeres}, {Marzari}, and {Rossi}]{sche04}
{Scheeres}, D.~J., {Marzari}, F., {Rossi}, A., 2004.
\newblock {Evolution of NEO rotation rates due to close encounters with Earth
  and Venus}.
\newblock Icarus~170, 312--323.

\bibitem[{Scheeres} et~al.(2006){Scheeres}, {Fahnestock}, {Ostro}, {Margot},
  {Benner}, {Broschart}, {Bellerose}, {Giorgini}, {Nolan}, {Magri}, {Pravec},
  {Scheirich}, {Rose}, {Jurgens}, {De Jong}, and {Suzuki}]{sche06s}
{Scheeres}, D.~J., and 15 colleagues, 2006.
\newblock {Dynamical configuration of binary near-Earth asteroid (66391) 1999
  KW4}.
\newblock Science~314, 1280--1283.

\bibitem[{Schellart}(2000){Schellart}]{schellart00}
{Schellart}, W.~P., 2000.
\newblock {Shear test results for cohesion and friction coefficients for
  different granular materials: Scaling implications for their usage in
  analogue modelling}.
\newblock Tectonophys.~324, 1--16.

\bibitem[{Sharma}(2009){Sharma}]{shar09}
{Sharma}, I., 2009.
\newblock {The equilibrium of rubble-pile satellites: The Darwin and Roche
  ellipsoids for gravitationally held granular aggregates}.
\newblock Icarus~200, 636--654.

\bibitem[{Sharma}(2010){Sharma}]{shar10}
{Sharma}, I., 2010.
\newblock {Equilibrium shapes of rubble-pile binaries: The Darwin ellipsoids
  for gravitationally held granular aggregates}.
\newblock Icarus~205, 638--657.

\bibitem[{Sharma} et~al.(2006){Sharma}, {Jenkins}, and {Burns}]{shar06}
{Sharma}, I., {Jenkins}, J.~T., {Burns}, J.~A., 2006.
\newblock {Tidal encounters of ellipsoidal granular asteroids with planets}.
\newblock Icarus~183, 312--330.

\bibitem[{Smith} and {Born}(1976){Smith} and {Born}]{smit76}
{Smith}, J.~C., {Born}, G.~H., 1976.
\newblock {Secular acceleration of Phobos and Q of Mars}.
\newblock Icarus~27, 51--53.

\bibitem[{Sridhar} and {Tremaine}(1992){Sridhar} and {Tremaine}]{srid92}
{Sridhar}, S., {Tremaine}, S., 1992.
\newblock {Tidal disruption of viscous bodies}.
\newblock Icarus~95, 86--99.

\bibitem[{Szeto}(1983){Szeto}]{szet83}
{Szeto}, A.~M.~K., 1983.
\newblock {Orbital evolution and origin of the martian satellites}.
\newblock Icarus~55, 133--168.

\bibitem[{Taylor} and {Margot}(2007){Taylor} and {Margot}]{tayl07dpstides}
{Taylor}, P.~A., {Margot}, J.~L., 2007.
\newblock {Tidal evolution of solar system binaries}.
\newblock Bull. Am. Astron. Soc.~39, 439.

\bibitem[{Taylor} and {Margot}(2010){Taylor} and {Margot}]{tayl10material}
{Taylor}, P.~A., {Margot}, J.~L., 2010.
\newblock {Binary asteroid systems: Tidal end states and estimates of material
  properties}.
\newblock submitted.

\bibitem[{Vokrouhlick{\'y}} and {{\v C}apek}(2002){Vokrouhlick{\'y}} and {{\v
  C}apek}]{vokr02}
{Vokrouhlick{\'y}}, D., {{\v C}apek}, D., 2002.
\newblock {YORP-induced long-term evolution of the spin state of small
  asteroids and meteoroids: Rubincam's approximation}.
\newblock Icarus~159, 449--467.

\bibitem[{Walsh} and {Richardson}(2006){Walsh} and {Richardson}]{wals06}
{Walsh}, K.~J., {Richardson}, D.~C., 2006.
\newblock {Binary near-Earth asteroid formation: Rubble pile model of tidal
  disruptions}.
\newblock Icarus~180, 201--216.

\bibitem[{Walsh} et~al.(2008){Walsh}, {Richardson}, and {Michel}]{wals08yorp}
{Walsh}, K.~J., {Richardson}, D.~C., {Michel}, P., 2008.
\newblock {Rotational breakup as the origin of small binary asteroids}.
\newblock Nature~454, 188--191.

\bibitem[Weidenschilling et~al.(1989)Weidenschilling, Paolicchi, and
  {Zappal\`a}]{weid89}
Weidenschilling, S.~J., Paolicchi, P., {Zappal\`a}, V., 1989.
\newblock {Do asteroids have satellites?}
\newblock In: Binzel, R.~P., Gehrels, T., Matthews, M.~S. (Eds.), Asteroids II.
  Univ. of Arizona Press, Tucson, pp.\  643--660.

\end{thebibliography}


\pagebreak

\renewcommand{\arraystretch}{0.5}

\begin{table}[!h]
\begin{center}
\begin{tabular}{ccc}
\hline\noalign{\smallskip}
$\ell$ & $a/R_{\rm p}$ & Legendre Polynomial, P$_{\ell}$$\left(\cos\psi\right)$ \\
\noalign{\smallskip}\hline\noalign{\smallskip}
 2 & 4.64 & $\frac{1}{4}\left(1\,+\,3\,\cos\,2\psi\right)$ \\
   &      & \\
 3 & 3.16 & $\frac{1}{8}\left(3\,\cos\,\psi\,+\,5\,\cos\,3\psi\right)$ \\
   &      & \\
 4 & 2.51 & $\frac{1}{64}\left(9\,+\,20\,\cos\,2\psi\,+\,35\,\cos\,4\psi\right)$ \\
   &      & \\
 5 & 2.15 & $\frac{1}{128}\left(30\,\cos\,\psi\,+\,35\,\cos\,3\psi\,+\,63\,\cos\,5\psi\right)$ \\
   &      & \\
 6 & 1.93 & $\frac{1}{512}\left(50\,+\,105\,\cos\,2\psi\,+\,126\,\cos\,4\psi\,+\,231\,\cos\,6\psi\right)$ \\
\noalign{\smallskip}\hline
\end{tabular}
\end{center}
\caption{Order $\ell$ of Legendre polynomials necessary in the expansion of the gravitational potential
(\ref{eq:Vlegendre}) of a binary system (with $\psi=0$) to accurately reproduce the full potential 
(\ref{eq:Vfull}) to within 1\% at separations less than $a/R_{\rm p}\simeq 5$.  If $a/R_{\rm p}$ is greater
than the value listed, expansion to the corresponding order $\ell$ suffices.  Recall that the fluid 
Roche limit is $a/R_{\rm p}=2.46$ (see Section \ref{sec:roche}).  Also note the Legendre polynomials are 
given in terms of $\cos~m\psi$, where $m$ is an integer, rather than the more common form of $\cos^{m}\psi$.}
\label{tab:legendre}
\end{table}

\thispagestyle{empty}

\clearpage

\renewcommand{\arraystretch}{0.5}

\begin{table}[!h]
\begin{center}
\begin{tabular}{cccccc}
\hline\noalign{\smallskip}
$\dot{a}_{\ell}/\dot{a}$ & $a/R_{\rm p}=1.93$ & $a/R_{\rm p}=2.15$ & $a/R_{\rm p}=2.51$ & $a/R_{\rm p}=3.16$ & $a/R_{\rm p}=4.64$ \\
\noalign{\smallskip}\hline\noalign{\smallskip}
 $\dot{a}_{2}/\dot{a}$  & 76.25\% & 80.93\% & 86.08\% & 91.27\% & 95.97\% \\
                        &        &        &        &        &        \\
 $\dot{a}_{3}/\dot{a}$  & 17.68\% & 15.12\% & 11.80\% & 7.89\% & 3.85\% \\
                        &        &        &        &        &        \\
 $\dot{a}_{4}/\dot{a}$  & 4.55\% & 3.14\% & 1.80\% & 0.76\% & 0.17\% \\
                        &        &        &        &        &        \\
 $\dot{a}_{5}/\dot{a}$  & 1.20\% & 0.67\% & 0.28\% & 0.07\% & 0.01\% \\
                        &        &        &        &        &        \\
 $\dot{a}_{6}/\dot{a}$  & 0.32\% & 0.14\% & 0.04\% & 0.01\% & -- \\
\noalign{\smallskip}\hline
\end{tabular}
\end{center}
\caption{Maximum contributions by the successive orders $\ell$ that alter the semimajor axis of 
the mutual orbit in (\ref{eq:adotboth}) at the separations listed in Table~\ref{tab:legendre}.  
The strengths of the contributions depend on the size ratio of the components with systems having 
negligibly small secondaries or equal-size components having the strongest contributions from higher 
order terms, which are shown here.  Having a synchronous secondary ($\omega_{\rm s}=n$) also has the
same effect on $\dot{a}_{\ell}/\dot{a}$.  It is assumed the components have similar $\mu Q$ parameters 
and the effect of each component's tides on the semimajor axis are additive.}
\label{tab:astrength}
\end{table}

\thispagestyle{empty}


\pagebreak

\addtolength{\voffset}{-0.5in}

\thispagestyle{empty}

\clearpage 

\thispagestyle{empty}

\clearpage 

\thispagestyle{empty}

\clearpage 

\thispagestyle{empty}

\clearpage 

\thispagestyle{empty}

\clearpage 

\thispagestyle{empty}

\clearpage 

\thispagestyle{empty}

\clearpage 

\thispagestyle{empty}

\end{document}